\documentclass[aps, prd, amsmath, floats, floatfix, twocolumn, nofootinbib, superscriptaddress, showpacs]{revtex4-1}
\usepackage{graphicx}
\usepackage{color}
\usepackage{latexsym}

\newcommand{\be}{\begin{eqnarray}}
\newcommand{\ee}{\end{eqnarray}}

\begin{document}

\title{Blandford-Znajek mechanism in black holes in alternative theories of gravity}

\author{Guancheng Pei}
\affiliation{Center for Field Theory and Particle Physics and Department of Physics, Fudan University, 200433 Shanghai, China}

\author{Sourabh Nampalliwar}
\affiliation{Center for Field Theory and Particle Physics and Department of Physics, Fudan University, 200433 Shanghai, China}

\author{Cosimo Bambi}
\email[Corresponding author: ]{bambi@fudan.edu.cn}
\affiliation{Center for Field Theory and Particle Physics and Department of Physics, Fudan University, 200433 Shanghai, China}
\affiliation{Theoretical Astrophysics, Eberhard-Karls Universit\"at T\"ubingen, 72076 T\"ubingen, Germany}

\author{Matthew J. Middleton}
\affiliation{Institute of Astronomy, University of Cambridge, Madingley Road, Cambridge CB3 0HA, UK}

\date{\today}

\begin{abstract}
According to the Blandford-Znajek mechanism, black hole jets are powered by the rotational energy of the compact object. In this work, we consider the possibility that the metric around black holes may not be described by the Kerr solution and we study how this changes the Blandford-Znajek model. If the Blandford-Znajek mechanism is responsible for the formation of jets, the estimate of the jet power in combination with another measurement can test the nature of black hole candidates and constrain possible deviations from the Kerr solution. However, this approach might become competitive with respect to other techniques only when it will be possible to have measurements much more precise than those available today.
\end{abstract}

\maketitle


\section{Introduction}
Accretion disks around black holes are an ubiquitous feature in the Universe.  Relativistic jets and outflows are a common phenomena associated with such accreting objects. These jets may carry a large fraction of the accreting energy. In the case of black hole binaries, we observe two kinds of jets~\cite{jets}. {\it Steady jets} commonly appear when the source is in the hard state, over a wide range of accretion luminosities. {\it Transient jets} show up when the source switches from the hard to the soft state at high accretion rates, but there is also some evidence of transient jets when the source moves from the soft to the hard state at low accretion rates before returning to quiescence~\cite{rev-belloni}. The mechanism for the formation of jets is currently unknown. An appealing model is the Blandford-Znajek (BZ) mechanism~\cite{bz77}, in which jets are powered by the rotational energy of the compact object.

Possibilities of a correlation between jet power and black hole spin have been explored in literature but there is no consensus on such a correlation, mainly because it is difficult to estimate the jet power, the uncertainty is large, and currently there are just a few measurements. In Refs.~\cite{f1,f2}, the authors consider both steady and transient jets and spin measurements from the iron line and the continuum-fitting methods; see Ref.~\cite{rev-matt} for a review. Their plots do not show any correlation between jet power and black hole spin. The authors of Ref.~\cite{n1} find instead a correlation between the estimate of the power of transient jets and the black hole spin measurements (from the continuum-fitting method) in only those sources which meet the selection criteria of reaching their Eddington limit. The issue will presumably be solved when more data will be available~\cite{n2,n3}.

Since the rotational energy of a black hole depends on the metric of the spacetime, the BZ mechanism can potentially test the nature of black hole candidates~\cite{io1,io2}. Here we want to explore whether this technique can compete with other approaches discussed in the literature to test the Kerr paradigm~\cite{srev,sss2,kk1,kk2,p1,sss1,p2,p3,p4,sss3}, or what level of precision in the measurements is required for this technique to be applicable. In the present work, we start with the BZ mechanism as our model for the formation of transient jets in black hole binaries and we study this model in other spacetimes. We also compare the jet power with the spin measurements from the continuum-fitting method to investigate whether current observations can tell us something about the metric near black hole candidates. We find that current measurements are not yet precise enough to test the Kerr metric although as we discuss, future improvements may address this.

\section{Blandford-Znajek mechanism in the Kerr metric}

In the original BZ model, we have a Kerr black hole surrounded by a stationary, axisymmetric, force-free, magnetized plasma~\cite{bz77,bz1,bzj}. In the force-free approximation, the energy-momentum tensor is that of the electromagnetic field (that is, matter is ignored)
\be
T^{\mu\nu} = T^{\mu\nu}_{\rm em} =
F^{\mu\rho} F^\nu_\rho - \frac{1}{4} g^{\mu\nu} F^{\sigma\tau} F_{\sigma\tau} \, ,
\ee
where $F_{\mu\nu} = \partial_\mu A_\nu - \partial_\nu A_\mu$ is the Faraday tensor and $A_\mu$ is the vector potential. The equations of motion are
\be\label{eq-motion}
\nabla_\mu T^{\mu\nu}_{\rm em} = 0 \, .
\ee

Assuming the force free condition, if $A_\mu$ is independent of the $t$ and $\phi$ coordinates, the Faraday tensor is~\cite{bz1}
\be
F_{\mu\nu} = \sqrt{-g}
\left( \begin{array}{cccc}
0 & - \omega B^\theta & \omega B^r & 0 \\
\omega B^\theta & 0 & B^\phi & - B^\theta \\
- \omega B^r & -B^\phi & 0 & B^r \\
0 & B^\theta & -B^r & 0
\end{array} \right) \, ,
\ee
where
\be
\omega = - \frac{\partial_\theta A_t}{\partial_\theta A_\phi}
= - \frac{\partial_r A_t}{\partial_r A_\phi}
\ee
is the rotational frequency of the electromagnetic field, $g$ is the determinant of the metric of the spacetime $g_{\mu\nu}$, $B^i = \tilde{F}^{it}$ is the magnetic field, and $\tilde{F}^{\mu\nu}$ is the dual of the Faraday tensor.

The total energy flux from the black hole is
\be\label{eq-pbz}
P_{\rm BZ} = - 2 \pi \int_0^\pi d\theta \, \sqrt{-g} \, T^r_t \, ,
\ee
where $T^r_t$ is the radial component of the Poynting vector and the integral is evaluated at some surface $r = const$. Eq.~(\ref{eq-pbz}) is independent of the choice of radial coordinate at which the integration is performed and of the shape of the magnetic field.

Eq.~(\ref{eq-pbz}) is usually calculated in Kerr-Schild coordinates, which are regular at the event horizon. With this choice, the line element of the Kerr metric reads
\be
ds^2 &=& - \left(1 - \frac{2Mr}{\Sigma}\right) dt^2
+ \left(1 + \frac{2Mr}{\Sigma} \right) dr^2
+ \Sigma \, d\theta^2
\nonumber\\
&&
+ \left(r^2 + a^2 + \frac{2Ma^2r \sin^2\theta}{\Sigma}\right) \sin^2\theta \, d\phi^2
\nonumber\\
&&
+ \frac{4Mr}{\Sigma} \, dt \, dr
- \frac{4Mar \sin^2\theta}{\Sigma} \, dt \, d\phi
\nonumber\\
&&
- 2 a \left(1 + \frac{2Mr}{\Sigma}\right) \sin^2\theta \, dr \, d\phi \, ,
\ee
where $\Sigma = r^2 + a^2\cos^2\theta$ and $a = J/M$ is the specific angular momentum. In what follows, we will also use the spin parameter $a_* = a/M$, which is dimensionless. The radial component of the Poynting vector is
\be\label{eq-poy}
T^r_t &=& 2 \left( B^r \right)^2 \omega r \left(\omega - \frac{a}{2 M r}\right) \sin^2\theta
\nonumber\\ &&
+ B^r B^\phi \omega \Delta \sin^2\theta \, ,
\ee
where $\Delta = r^2 - 2 M r + a^2$. At the event horizon $\Delta = 0$ and Eq.~(\ref{eq-poy}) becomes
\be
T^r_t = 2 \left( B^r \right)^2 \omega r_{\rm H}
\left(\omega - \Omega_{\rm H}\right) \sin^2\theta \, ,
\ee
where $r_{\rm H} = M + \sqrt{M^2 - a^2}$ is the radial coordinate of the event horizon and $\Omega_{\rm H} = a/(2Mr_{\rm H})$ is the angular velocity of the event horizon.

The evaluation of Eq.~(\ref{eq-pbz}) requires a solution to Eq.~(\ref{eq-motion}) in order to find $B^r$ and $\omega$. Unfortunately, this is non-trivial. The standard approach (see, e.g.,~\cite{bzj,bz2}) is thus to find an exact solution of Eq.~(\ref{eq-motion}) for the Schwarzschild spacetime and then consider an expansion in $a$ or $\Omega_{\rm H}$ to find the rotating solution perturbatively.

We follow the expansion in $\Omega_{\rm H}$ as performed by ~\cite{bz2}. At the leading order in $\Omega_{\rm H}$, the BZ formula for the jet power can be written as
\be\label{eq-pbz1}
P_{\rm BZ} = \frac{\kappa}{16 \pi} \Phi_{\rm B}^2 \Omega_{\rm H}^2 + O (\Omega_{\rm H}^4) \, ,
\ee
where $\kappa$ is a numerical constant which depends on the magnetic field configuration (for instance, $\kappa = 0.053$ for a split monopole geometry and 0.044 for a parabolic geometry), and $\Phi_{\rm B}$ is the magnetic flux threading the black hole horizon
\be
\Phi_{\rm B} = 2 \pi \int_0^\pi | B^r | \sqrt{-g} \, d\theta \, .
\ee
Including higher order terms, Eq.~(\ref{eq-pbz1}) becomes
\be
P_{\rm BZ} = \frac{\kappa}{16 \pi} \Phi^2 \Omega_{\rm H}^2 f(\Omega_{\rm H}^2)
\ee
for some function
\be
f(\Omega_{\rm H}^2) = 1 + c_1 \Omega_{\rm H}^2 + c_2 \Omega_{\rm H}^4 + ... \, ,
\ee
where $\{ c_i \}$ are certain numerical coefficients. Note that Eq.~(\ref{eq-pbz1}) is an expansion in $\Omega_{\rm H}$; an expression based on expansion in $a$ performs worse. Moreover, the expression in Eq.~(\ref{eq-pbz1}) works quite well even when $a_*$ is quite close to 1~\cite{bz2}. Therefore, for the rest of the paper, we only consider $P_{\rm BZ}$ up to $\Omega_{\rm H}^2$, neglecting terms $O (\Omega_{\rm H}^4)$.

\section{Blandford-Znajek mechanism in the Johannsen metric}

We now consider the BZ mechanism in the Johannsen metric~\cite{j} as a prototype of jets in alternative theories of gravity. While there are many non-Kerr metrics, we adopt the Johannsen one because: $i)$ it is an analytic metric without restrictions on the value of the spin parameter $a_*$ (analytic black hole solutions in alternative theories of gravity are typically known in the slow-rotation approximation), $ii)$ its black holes have a regular exterior (no closed time-like curves or naked singularities), and $iii)$ the angular velocity of the event horizon is well defined. In Boyer-Lindquist coordinates, the line element of the Johannsen metric reads
\be\label{eq-metric-j}
ds^2 &=&
- \frac{\left( \Delta - a^2 A_2^2 \sin^2\theta\right)
\tilde{\Sigma}}{\left[\left(r^2 + a^2\right) A_1 - a^2 A_2 \sin^2\theta\right]^2} \, dt^2
\nonumber\\ &&
- \frac{2a \left[\left(r^2 + a^2\right) A_1 A_2 - \Delta\right] \tilde{\Sigma} \sin^2\theta}{\left[\left(r^2 + a^2\right) A_1 - a^2 A_2 \sin^2\theta\right]^2} \, dt \, d\phi
\nonumber\\ &&
+ \frac{\left[\left(r^2 + a^2\right)^2 A_1^2 - a^2 \Delta \sin^2\theta\right] \tilde{\Sigma} \sin^2\theta}{\left[\left(r^2 + a^2\right) A_1 - a^2 A_2 \sin^2\theta\right]^2} \, d\phi^2
\nonumber\\ &&
+ \frac{\tilde{\Sigma}}{\Delta A_5} \, dr^2 + \tilde{\Sigma} \, d\theta^2 \, ,
\ee
where $\tilde{\Sigma} = \Sigma + f$, and $A_1$, $A_2$, $A_5$, and $f$ are some functions that are introduced to describe possible deviations from the Kerr solution. The Kerr metric is recovered for $A_1 = A_2 = A_5 = 1$ and $f = 0$. In their simplest forms, these functions are
\be
A_1 &=& 1 + \alpha_{13} \left(\frac{M}{r}\right)^3 \, , \\
A_2 &=& 1 + \alpha_{22} \left(\frac{M}{r}\right)^2 \, , \\
A_5 &=& 1 + \alpha_{52} \left(\frac{M}{r}\right)^2 \, , \\
f &=& \epsilon_3 \frac{M^3}{r} \, ,
\ee
where $\alpha_{13}$, $\alpha_{22}$, $\alpha_{52}$, and $\epsilon_3$ are the ``deformation parameters''. The position of the event horizon is given by
\be
\Delta = r^2 - 2Mr + a^2 = 0 \, ,
\ee
and therefore it is the same as the Kerr metric for the same $M$ and $a$. The angular frequency of the event horizon is
\be
\Omega_{\rm H} =
\left( - \frac{g_{t\phi}}{g_{\phi\phi}} \right)_{r = r_{\rm H}}
= \frac{a A_2}{2 M r_{\rm H} A_1}
= \frac{A_2}{A_1} \, \Omega_{\rm H}^{\rm Kerr} \, .
\ee

The simplest non-Kerr model in the Johannsen metric is one in which the only non-vanishing deformation parameter is $\alpha_{22}$, while the others are set to zero, i.e. $\alpha_{13} = \alpha_{52} = \epsilon_3 = 0$. In this case, non-rotating black holes are described by the Schwarzschild solution, independently of the value of $\alpha_{22}$. The solution of the non-rotating field is thus known. Eq.~(\ref{eq-pbz1}) then becomes
\be\label{eq-bz-a22}
P_{\rm BZ} = P_{\rm BZ}^{\rm Kerr} \left( 1 + \alpha_{22} \frac{M^2}{r^2} \right)^2
+ O (\Omega_{\rm H}^4) \, ,
\ee
where $P_{\rm BZ}^{\rm Kerr}$ is exactly the Kerr expression for the jet power~\cite{j}.

\begin{figure*}[t]
\begin{center}
\includegraphics[type=pdf,ext=.pdf,read=.pdf,width=7.5cm]{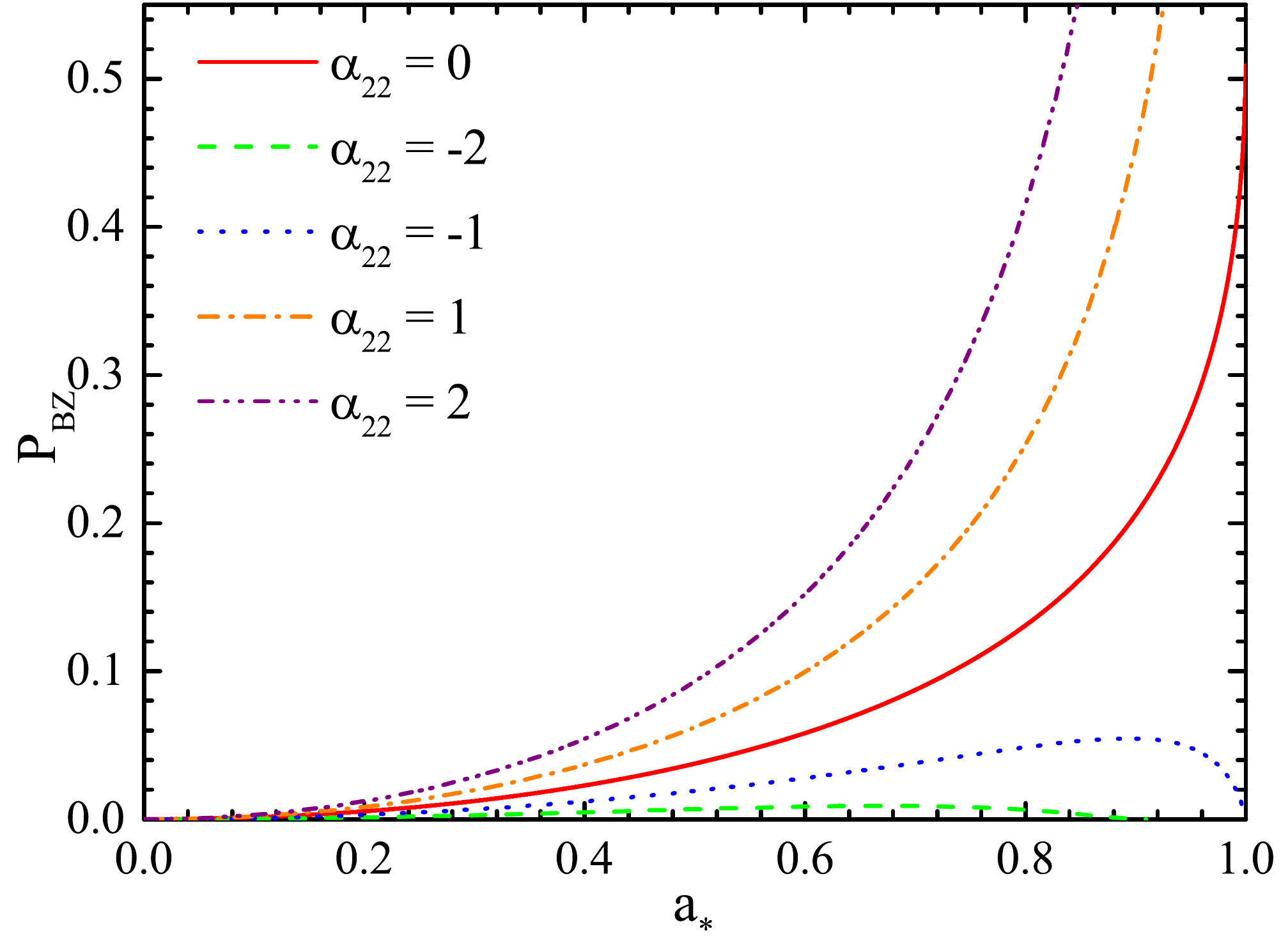}
\hspace{1.0cm}
\includegraphics[type=pdf,ext=.pdf,read=.pdf,width=7.5cm]{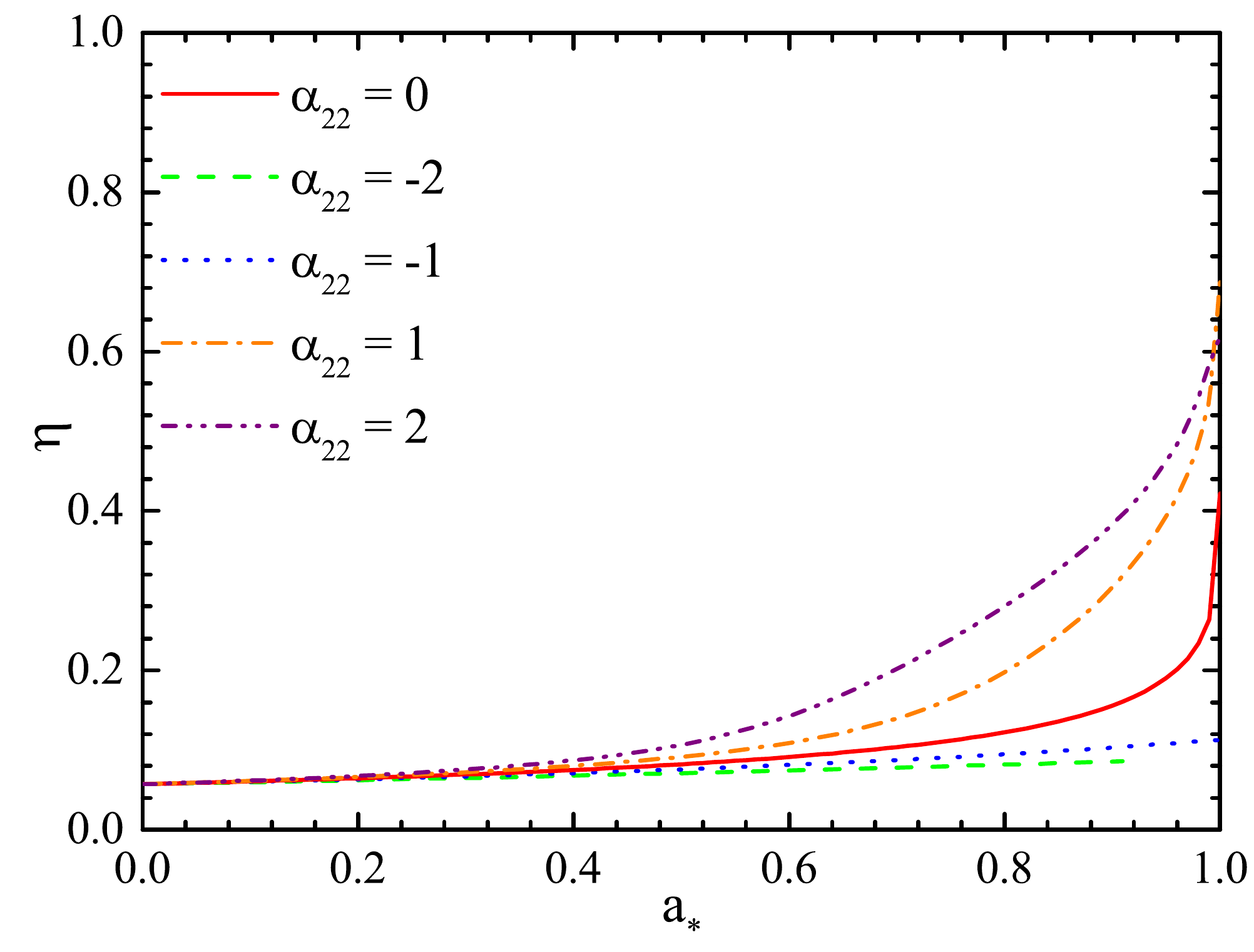} \\
\vspace{0.5cm}
\includegraphics[type=pdf,ext=.pdf,read=.pdf,width=7.5cm]{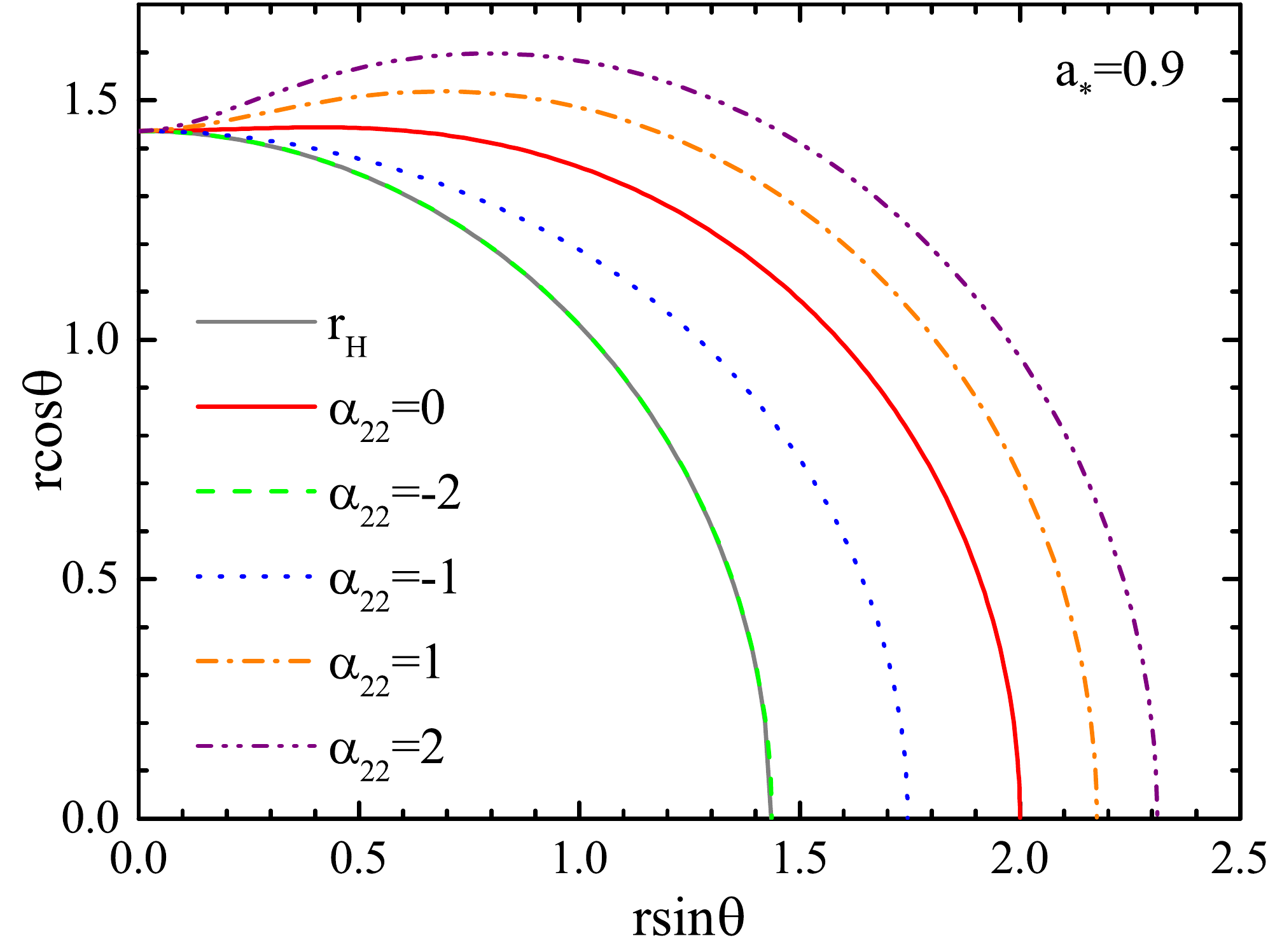}
\end{center}
\caption{Top panels: jet power (left panel) and Novikov-Thorne radiative efficiency $\eta = 1 - E_{\rm ISCO}$ (right panel) as a function of the spin parameter $a_*$ for Johannsen black holes with $\alpha_{22} = 0$, $\pm 1$, and $\pm 2$. Bottom panel: shape of the ergoregion of Johannsen black holes with $a_* = 0.9$ and $\alpha_{22} = 0$, $\pm 1$, and $\pm 2$ in the plane $(r\sin\theta, r\cos\theta)$, where $r$ and $\theta$ are the Boyer-Lindquist coordinates and the axes are in units $M=1$; each line describes the outer boundary of the ergoregion, i.e. the static limit given by Eq.~(\ref{eq-staticlimit}), and the grey solid line is for the event horizon. The other deformation parameters are always assumed to vanish. \label{f1}}
\end{figure*}

The top left panel in Fig.~\ref{f1} shows $P_{\rm BZ}$ as a function of $a_*$ for 5 different values of $\alpha_{22}$ assuming $\alpha_{13} = \alpha_{52} = \epsilon_3 = 0$. The red solid line for $\alpha_{22} = 0$ is the standard Kerr case. As $\alpha_{22}$ increases, $\Omega_{\rm H}$ and $P_{\rm BZ}$ increase. For $\alpha_{22} < 0$, we have the opposite case, and we also see that $P_{\rm BZ}$ is not a monotonic function of the spin $a_*$. For $\alpha_{22} = -2$, the jet power goes to zero for $a_* \approx 0.9$.

The behavior of $P_{\rm BZ}$ can be easily understood in terms of the size of the ergoregion, which is the exterior region in which $g_{tt} > 0$ and static observers (i.e. time-like and null-like geodesics) are not allowed. The outer boundary of the ergoregion is called the static limit and is defined by the largest root of $g_{tt} = 0$, namely
\be\label{eq-staticlimit}
\Delta - a^2 \left( 1 + \alpha_{22} \frac{M^2}{r^2} \right)^2 \sin^2\theta = 0 \, .
\ee
The bottom panel in Fig.~\ref{f1} shows the plane $(r\sin\theta, r\cos\theta)$ of the spacetime in units $M=1$. We set $a_* = 0.9$ and we plotted the static limit of the Johannsen black holes with $\alpha_{22} = 0$, $\pm 1$, and $\pm 2$. The event horizon is defined by $\Delta = 0$ and is shown by the grey line. For $a_* = 0.9$, the radius of the event horizon is $r_{\rm H} = 1.4359$~$M$. For $\alpha_{22} > 0$ ($< 0$), the ergoregion expands (contracts) with respect to the Kerr case (red solid line) and thus the jet power increases (decreases). Rotating Johannsen black holes have no ergoregion when the largest root of Eq.~(\ref{eq-staticlimit}) is smaller than the larger root of the equation $\Delta = 0$, namely when there is no static limit outside of the event horizon. For $a_* = 0.9$ and $\alpha_{22} = -2$, the radius of the static limit in the equatorial plane is $r_{\rm sl} = 1.4368$~$M$, and therefore the ergoregion is small but exists. The ergoregion completely disappears if $a_* \ge 0.914$. Without ergoregion, it is impossible to extract the rotational energy of the black hole. Our result is consistent, and complementary, with previous studies, in which it was shown that the existence of the ergoregion is the key-ingredient to make the BZ mechanism work~\cite{v3-c1,v3-c2,v3-c3}. Here we see that the BZ mechanism cannot work if the ergoregion does not exist, even if the black hole is rotating. As we detect jets from black hole binary systems, invoking the extraction of energy from the BZ mechanism automatically excludes such extreme deformation values.

If the non-vanishing deformation parameter is $\alpha_{13}$ and the others are set to zero, we do not know the value of the magnetic field, but we can still say that $P_{\rm BZ}$ must be proportional to the square of the angular velocity of the event horizon, neglecting terms $O (\Omega_{\rm H}^4)$, namely
\be\label{eq-bz-a13}
P_{\rm BZ} &=& k_{13} \Omega_{\rm H}^2 + O (\Omega_{\rm H}^4) \\
&=& k_{13} \left( 1 + \alpha_{13} \frac{M^3}{r_{\rm H}^3} \right)^{-2}
\frac{a^2}{4 M^2 r_{\rm H}^2} + O (\Omega_{\rm H}^4) \, ,
\ee
where $k_{13} = k_{13} (\alpha_{13})$ is an unknown parameter. In order to determine $k_{13}$, we should find the magnetic field in the non-rotating limit for any non-vanishing $\alpha_{13}$.

The deformation parameters $\alpha_{52}$ and $\epsilon_3$ do not enter the expression of $\Omega_{\rm H}$. If they do not vanish, they may only alter the constant of proportionality between $P_{\rm BZ}$ and $\Omega_{\rm H}^2$. In such a case, the expression of the jet power reads, respectively for non-vanishing $\alpha_{52}$ and $\epsilon_3$,
\be
P_{\rm BZ} &=& k_{52} \frac{a^2}{4 M^2 r_{\rm H}^2} + O (\Omega_{\rm H}^4) \, , \\
P_{\rm BZ} &=& k_{3} \frac{a^2}{4 M^2 r_{\rm H}^2} + O (\Omega_{\rm H}^4) \, .
\ee
As in Eq.~(\ref{eq-bz-a13}), $k_{52} = k_{52}(\alpha_{52})$ and $k_{3} = k_{3}(\epsilon_3)$ should be calculated from the magnetic field solution in the non-rotating spacetime.

\section{Comparison with observations}

Since jet power does depend on the metric of the spacetime near the horizon, it is natural to ask whether there are some observational features in jet power related to the nature of black holes. Unfortunately, jet power also depends on the magnetic field strength which is not easy to unambiguously estimate (although polarimetric measures may assist in this regard). Presently, we can only test formulas like Eq.~(\ref{eq-bz-a13}).

In this section, we consider 6 black hole binaries for which we have both an estimate of the power of their transient jets and a measurement of the black hole spin via the continuum-fitting method. We have the 4 objects discussed in Ref.~\cite{n1}: GRS1915+105, GROJ1655-40, XTEJ1550-564, and A0620-00. The fifth source is H1743-322, which is discussed in Ref.~\cite{n2}. The sixth source is GRS 1124-683: the estimate of the jet power is reported in Ref.~\cite{n2}, while the spin has been measured via the continuum-fitting method only very recently in~\cite{lijun}. There would be a seventh source, the microquasar in M31, but there is only an upper bound for the value of the spin this is not useful for our purpose~\cite{mm31}.

For the estimate of the jet power, we follow the approach of Ref.~\cite{n1}, which is determined from the monochromatic flux density at 5 GHz, and is corrected for the distance to the source, de-boosted for two assumed values of the bulk Lorentz factor $\Gamma=2$ and 5~\cite{n2}, and divided by the black hole mass to remove any dependence. The values of the emitted flux density, $S_{\nu,0}$, are reported in Ref.~\cite{n2}, listed here in Tab.~\ref{ch03tab01} and the jet powers (as derived in~\cite{mm31}) are provided in Tab.~\ref{tab2} for convenience.

\begin{table*}[htbp]
\makebox[\textwidth][c]{
\begin{tabular*}{0.83\paperwidth}{@{\extracolsep{\fill}}ccccccccc}
  \hline\hline
  BH Binary & $a_*$ & $\eta$ & ${M(\mathrm{M_\odot})}$ & ${D(\mathrm{kpc})}$ & $i^\circ$ & ${(S_{\nu,0})_{\mathrm{max,5GHz}}(\mathrm{Jy})}$ & Reference \\ \hline
  A0620-00 & $0.12\pm{0.19}$ & $0.061^{+0.009}_{-0.007}$ & $6.61\pm{0.25}$ & $1.06\pm{0.12}$ & $51.0\pm0.9$ & 0.203 & \cite{ch03ref22,ch03ref23,ch03ref24} \\
  H1743-322 & $0.2\pm{0.3}$ & $0.065^{+0.017}_{-0.011}$ & $8.0$ & $8.5\pm{0.8}$ & $75.0\pm3.0$ & 0.0346 & \cite{n2} \\
  XTE J1550-564 & $0.34\pm024$ & $0.072^{+0.017}_{-0.011}$ & $9.10\pm{0.61}$ & $4.38\pm{0.5}$ & $74.7\pm3.8$ & 0.265 & \cite{ch03ref23,ch03ref25,ch03ref26} \\
  GRS 1124-683 & $0.63^{+0.16}_{-0.19}$ & $0.095^{+0.025}_{-0.017}$ & $11.0^{+2.1}_{-1.4}$ & $4.95^{+0.69}_{-0.65}$ & $50.5\pm6.5$ & 0.45 & \cite{n2,lijun} \\
  GRO J1655-40 & $0.7\pm{0.1}$ & $0.104^{+0.018}_{-0.013}$ & $6.30\pm{0.27}$ & $3.2\pm{0.5}$ & $70.2\pm1.9$ & 2.42 & \cite{ch03ref23,ch03ref26,ch03ref27,ch03ref28,ch03ref29} \\
  GRS 1915+105 & $0.975, a_*>0.95$ & $0.224,\eta>0.190$ & $12.4^{+1.7}_{-1.9}$ & $8.6^{+2.0}_{-1.6}$ & $60.0\pm5.0$ & 0.912 & \cite{ch03ref23,ch03ref30,ch03ref31,ch03ref32,refReid} \\
  \hline
\end{tabular*}
}
\caption{Parameters of transient black hole binaries.}
\label{ch03tab01}
\end{table*}

The measurement of the spin is a more subtle point, because it depends on the choice of the spacetime~\cite{cfmr}. In the literature, they are reported assuming the Kerr metric. The correct approach would be to repeat the data analysis of these 6~sources for the Johannsen metric. A simpler way would be to proceed as in Refs.~\cite{a-jp,a-cpr} and perform a simplified analysis assuming that the actual data are equal to the theoretical spectrum of a Kerr black hole with the spin equal to the value measured by the continuum-fitting method. Such a simplified approach is possible because the shape of the spectrum is eventually very simple, just a multi-color blackbody spectrum without specific features. However, the calculations require still quite a long time.

Here, we adopt an even more simple approach. It relies on the fact that the continuum-fitting method eventually measures something close to the radiative efficiency of the Novikov-Thorne model, i.e.
\be\label{eq-eta-eta}
\eta = 1 - E_{\rm ISCO} \, ,
\ee
where $E_{\rm ISCO}$ is the specific energy of a test-particle at the radius of the innermost stable circular orbit (ISCO). Such an approximation works better when the ISCO radius is larger and the inclination angle of the disk smaller, namely when relativistic effects are weaker, but it is not too bad even in the other cases (see e.g. the discussion in~\cite{a-jp}) and it has been repeatedly used in the past~\cite{io1,io2,qpo}. The rescaling works as follows. We take the spin measurement obtained within the Kerr metric, we evaluate $\eta$ via Eq.~(\ref{eq-eta-eta}), we find the spin parameter for which the Johannsen black hole under consideration has the same value of $\eta$. $\eta$ as a function of $a_*$ for different values of $\alpha_{22}$ is reported in the top right panel in Fig.~\ref{f1}.

Let us note that $\eta$ depends on $g_{tt}$, $g_{t\phi}$, and $g_{\phi\phi}$ (see e.g. Appendix~B in Ref.~\cite{enrico}). The spin measurements via the continuum-fitting methods are thus altered by $\alpha_{13}$, $\alpha_{22}$, and $\epsilon_3$. In what follows, we will not consider $\alpha_{52}$ any more, because it affects neither the angular frequency at the event horizon (at least if not in some extreme cases with the location of $r_{\rm H}$, when $A_5$ vanishes before $\Delta$) nor the ISCO radius and thus the spin measurement from the continuum-fitting technique.
\begin{table}[htbp]
  \centering
  \begin{tabular}{ccc}
  \hline\hline
  BH Binaries & $\Gamma=2,P_{\mathrm{jet}}$ & $\Gamma=5, P_{\mathrm{jet}}$ \\ \hline
  A0620-00 & 0.13 & 1.6 \\
  H1743-322 & 7.0 & 140 \\
  XTE J1550-564 & 11 & 180 \\
  GRS 1124-683 & 3.9 & 390 \\
  GRO J1655-40 & 70 & 1600 \\
  GRS 1915+105 & 42 & 660 \\
  \hline\hline
\end{tabular}
  \caption{Jet power proxy values in units of kpc$^2$~GHz~Jy~$M_\odot^{-1}$.}\label{tab2}
\end{table}

With this approach, we have reevaluated the nature of a putative correlation between jet power and spin in these 6 black hole binaries assuming the Johannsen metric, with one non-vanishing deformation parameter at a time. We have fitted the results with the formula
\be\label{eq-pbzpbz}
P_{\rm BZ} = k \Omega_{\rm H}^2 \, ,
\ee
where we let $k$ be a free parameter, to be determined by minimizing the error function $S$ defined below. Eq.~(\ref{eq-pbzpbz}) thus has two free parameters, namely $k$ and the non-vanishing deformation parameter under investigation.

The fit has been done by evaluating the function
\be
S (i,k) &=& \sum_j \frac{\left[ P_{{\rm jet},j} - P_{{\rm BZ},j}(i,k) \right]^2}{\sigma^2_{P,j}}
\nonumber\\
&& + \sum_j \frac{\left[ \Omega_{H,j} - \Omega_{H,j}^{\rm th}(i,k) \right]^2}{\sigma^2_{\Omega_H,j}} \, ,
\ee
where $i = \alpha_{13}$, $\alpha_{22}$, or $\epsilon_3$ and $k$ is the constant of proportionality in Eq.~(\ref{eq-pbzpbz}). $P_{{\rm jet},j}$ is the observational value, $P_{{\rm BZ},j}(i,k)$ is the theoretical prediction. The same distinction is for $\Omega_{H,j}$ and $\Omega_{H,j}^{\rm th}(i,k)$. $\sigma_{P,j}$ and $\sigma_{\Omega_H,j}$ are, respectively, the uncertainties on the estimate of the jet power and of the spin. $j = 1,...,6$ is the index for the source.

\begin{figure*}[t]
\begin{center}
\includegraphics[type=pdf,ext=.pdf,read=.pdf,width=7.5cm]{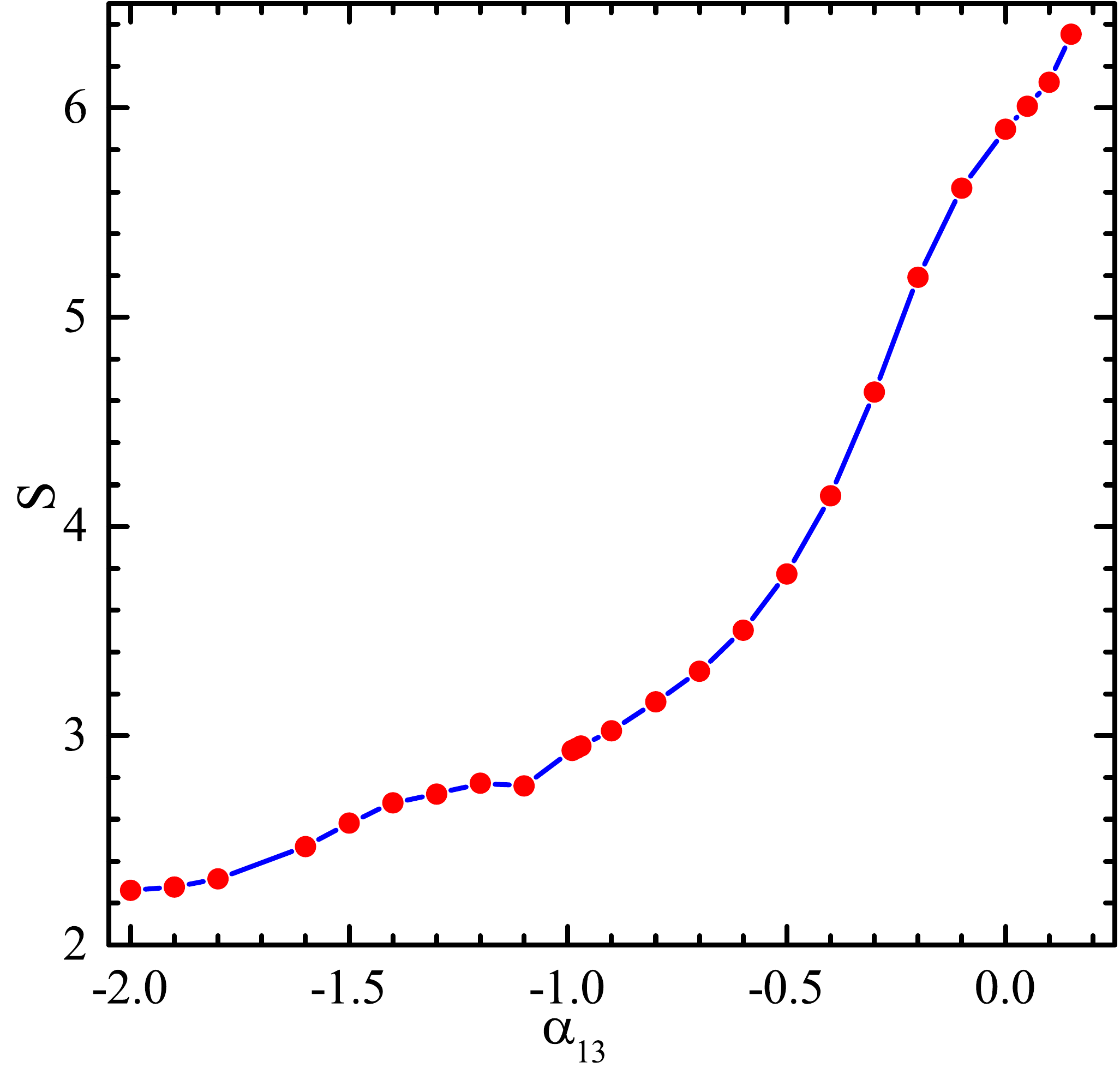}
\hspace{1.0cm}
\includegraphics[type=pdf,ext=.pdf,read=.pdf,width=7.5cm]{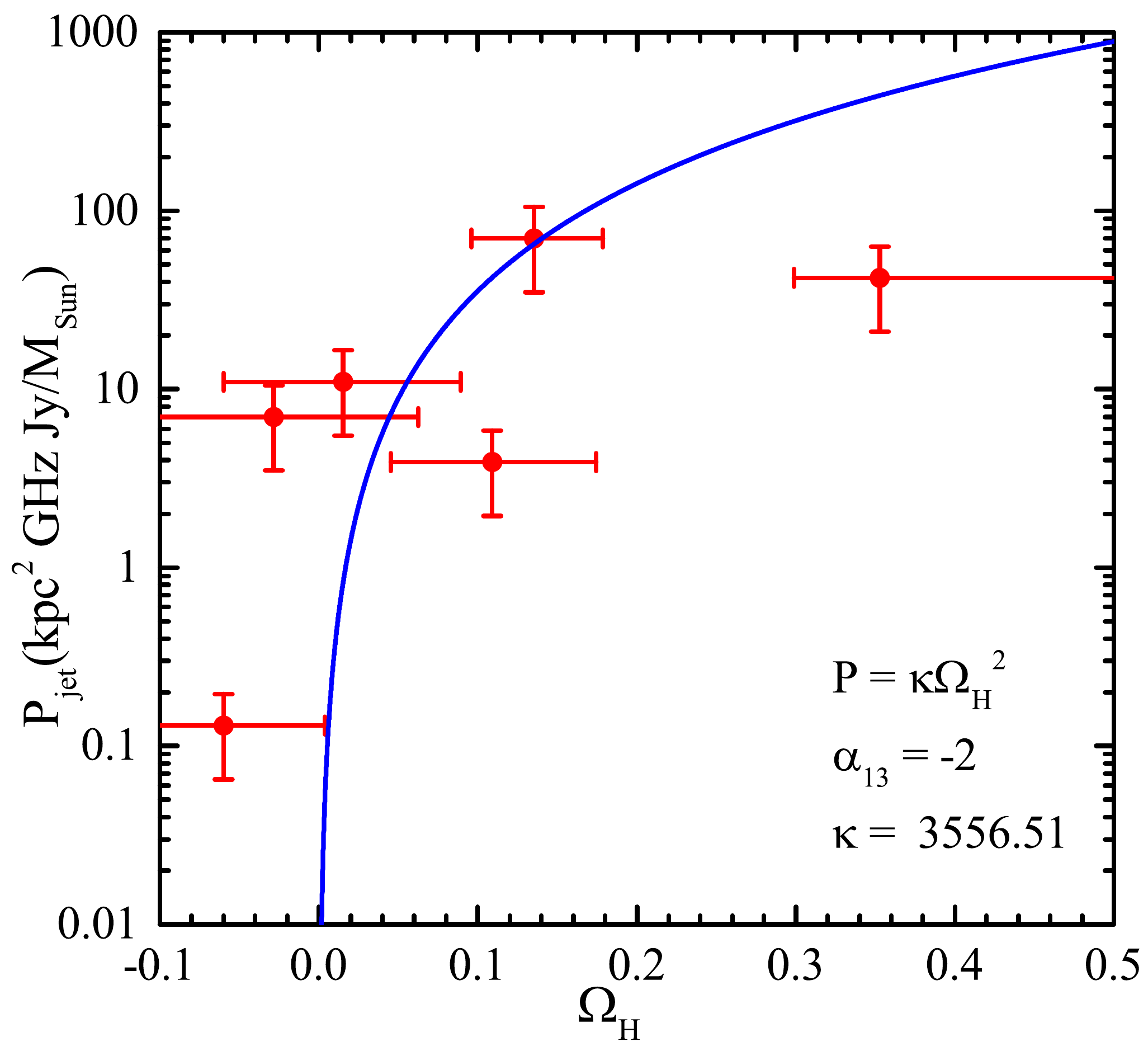} \\
\vspace{0.5cm}
\includegraphics[type=pdf,ext=.pdf,read=.pdf,width=7.5cm]{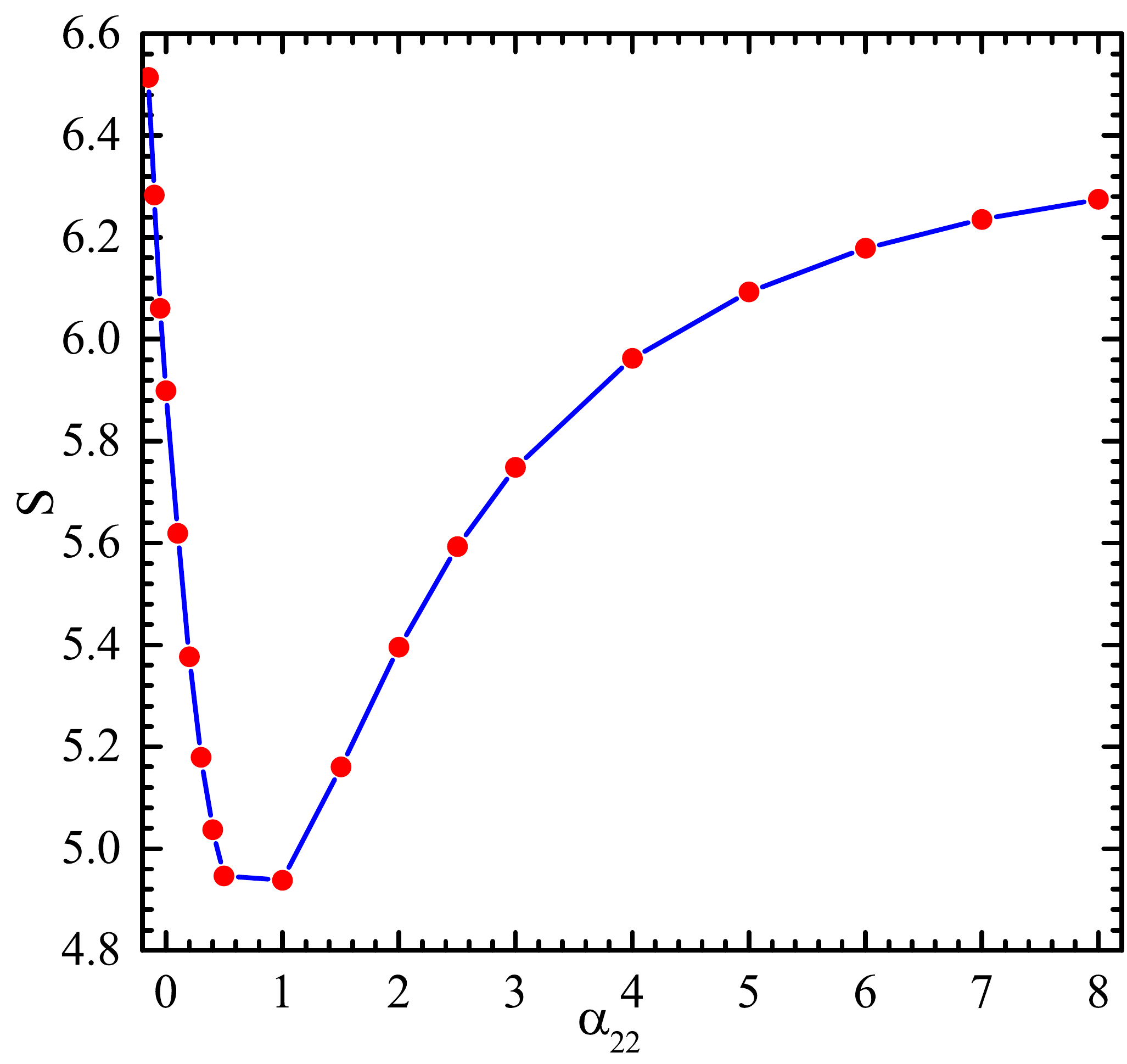}
\hspace{1.0cm}
\includegraphics[type=pdf,ext=.pdf,read=.pdf,width=7.5cm]{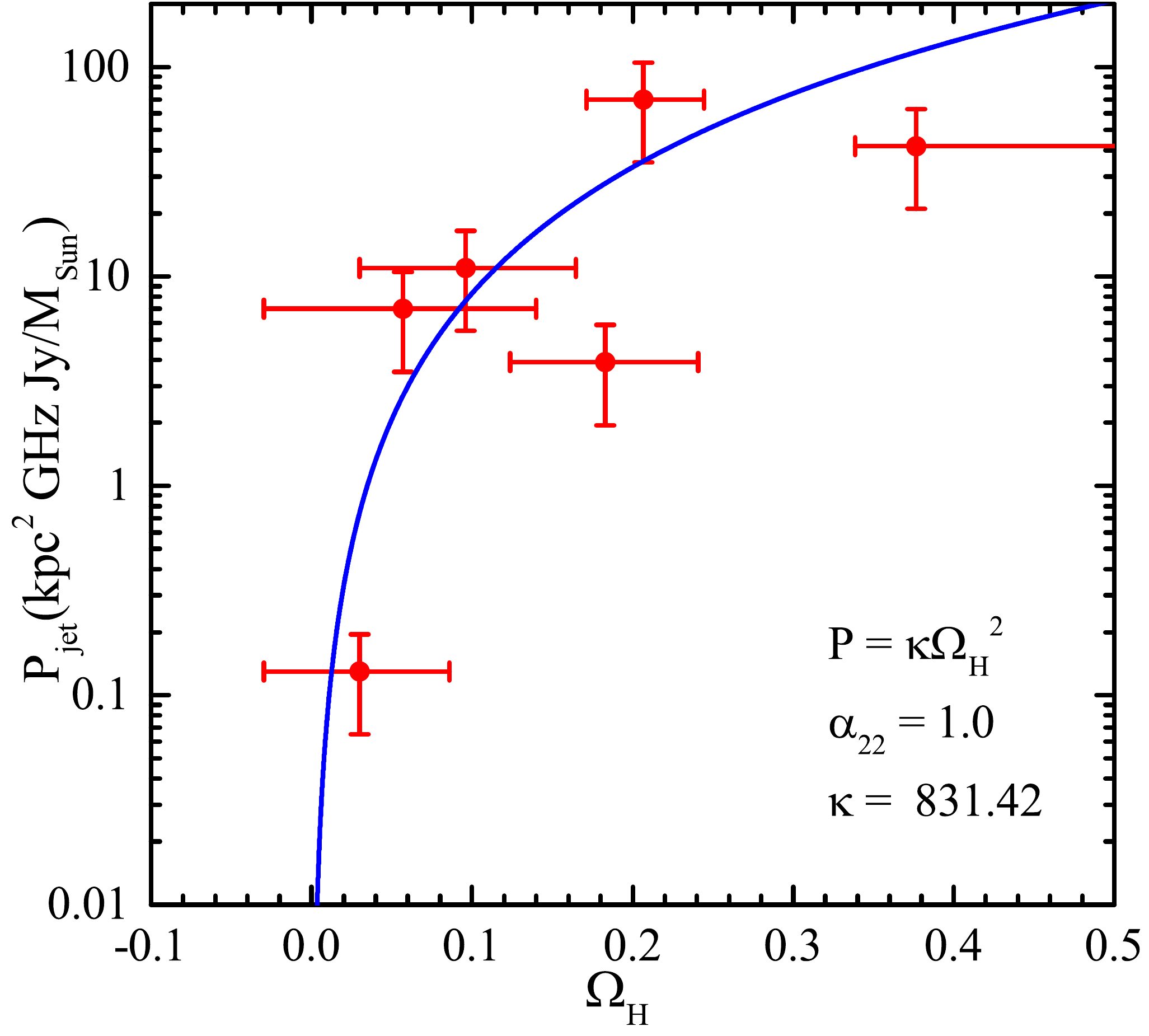} \\
\vspace{0.5cm}
\includegraphics[type=pdf,ext=.pdf,read=.pdf,width=7.5cm]{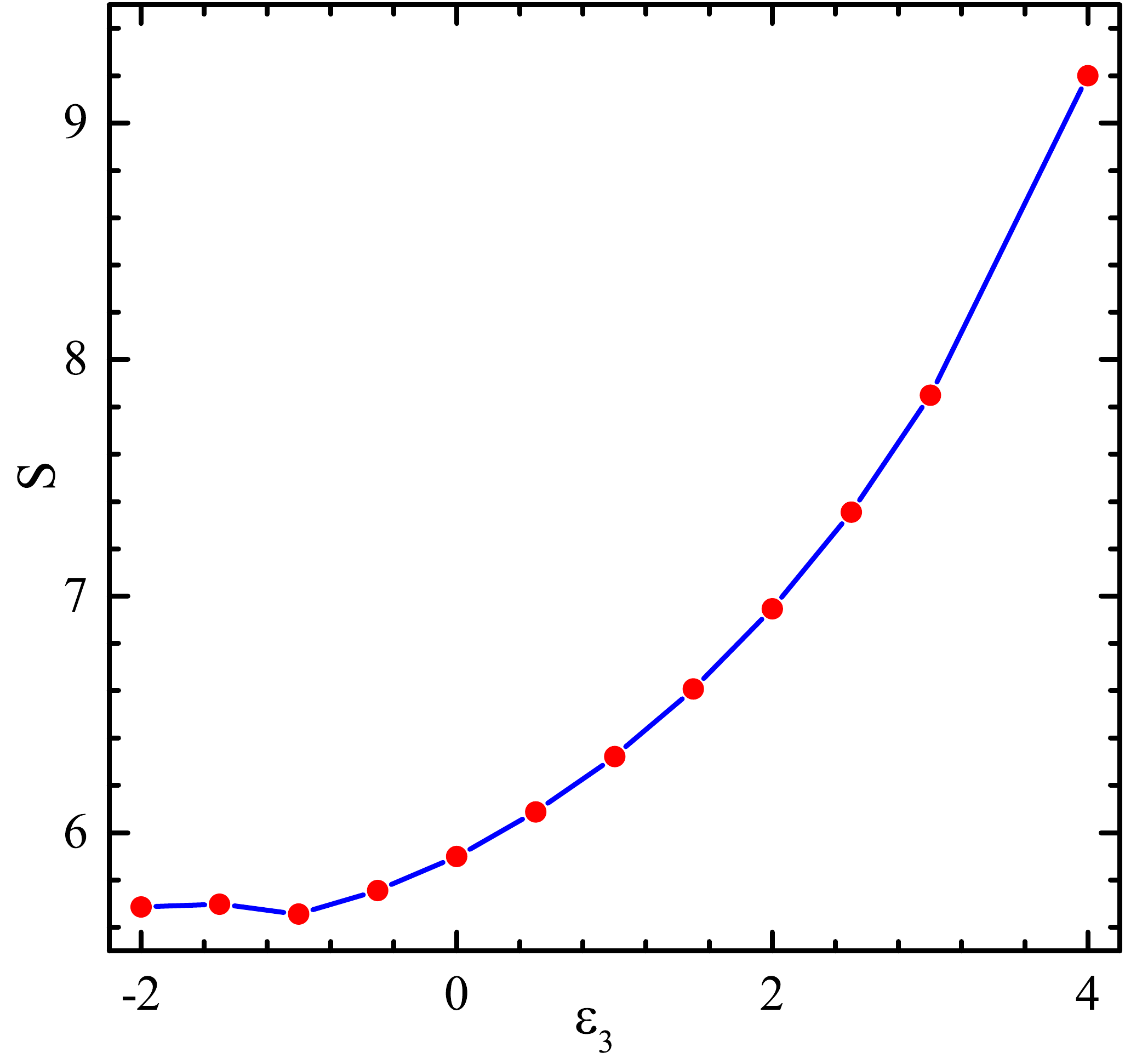}
\hspace{1.0cm}
\includegraphics[type=pdf,ext=.pdf,read=.pdf,width=7.5cm]{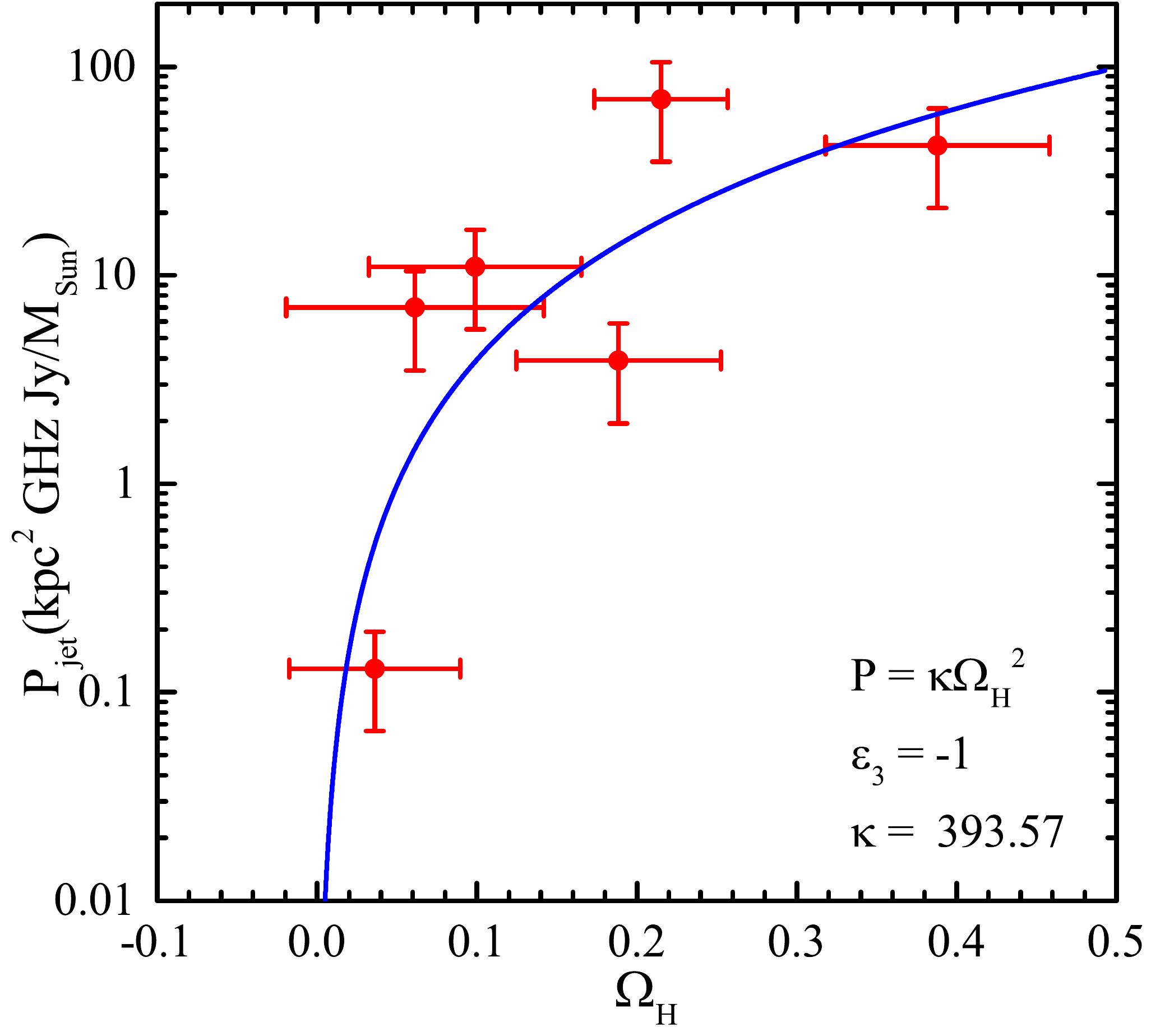} \\
\end{center}
\caption{Left panels: $S$ function of the 6 black hole binaries minimised over $k$ as a function of the deformation parameters assuming that all jets are produced with $\Gamma = 2$: $\alpha_{13}$ (top panel), $\alpha_{22}$ (middle panel), and $\epsilon_3$ (bottom panel). Right panels: Fitting plot of the 6 black hole binaries for the deformation parameters that minimise the $S$ function: $\alpha_{13}$ = $-2$ (top panel), $\alpha_{22}$ = $1$ (middle panel), and $\epsilon_3$ = $-1$ (bottom panel). \label{f2G2}}
\end{figure*}

\begin{figure*}[t]
\begin{center}
\includegraphics[type=pdf,ext=.pdf,read=.pdf,width=7.5cm]{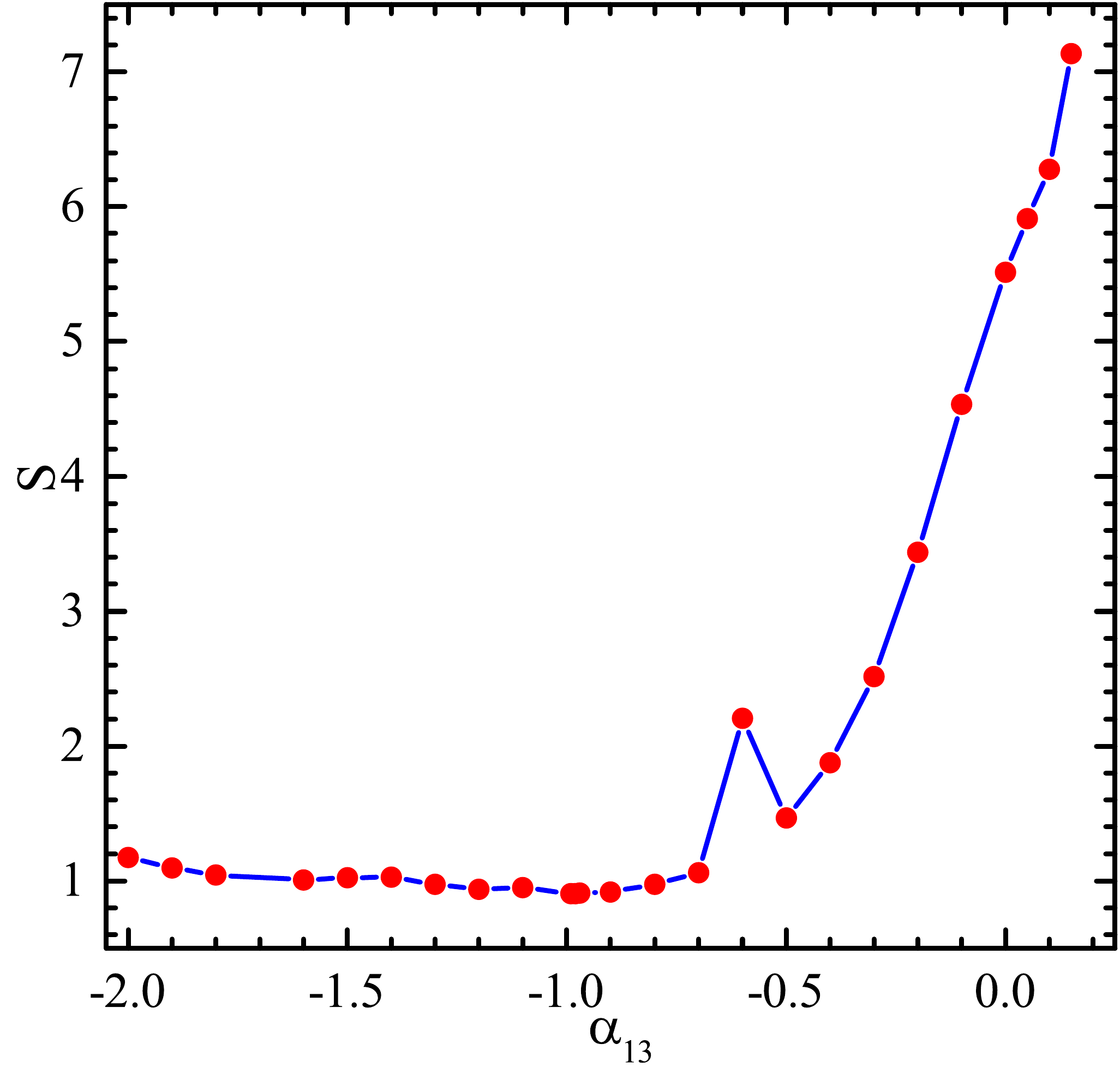}
\hspace{1.0cm}
\includegraphics[type=pdf,ext=.pdf,read=.pdf,width=7.5cm]{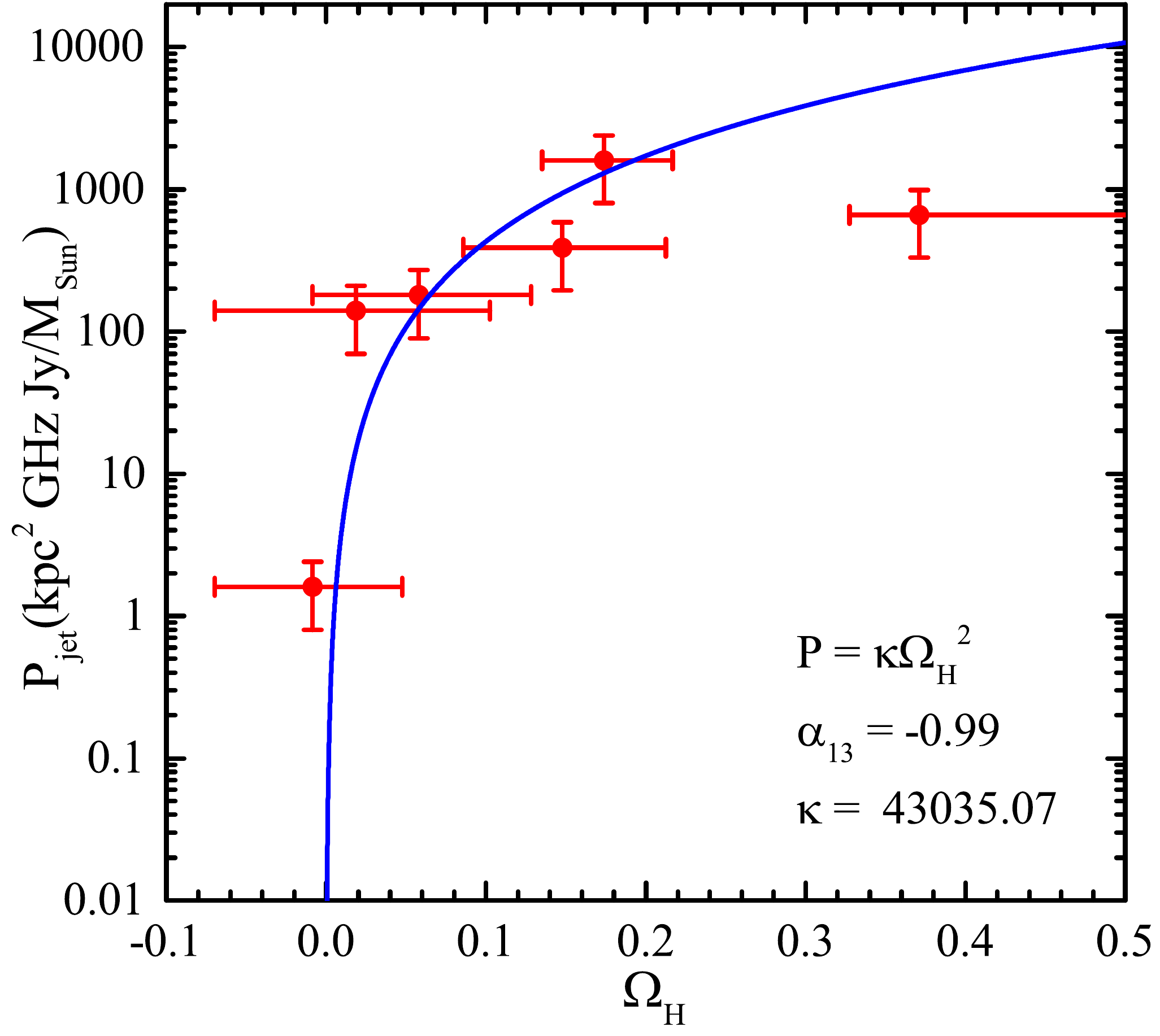} \\
\vspace{0.5cm}
\includegraphics[type=pdf,ext=.pdf,read=.pdf,width=7.5cm]{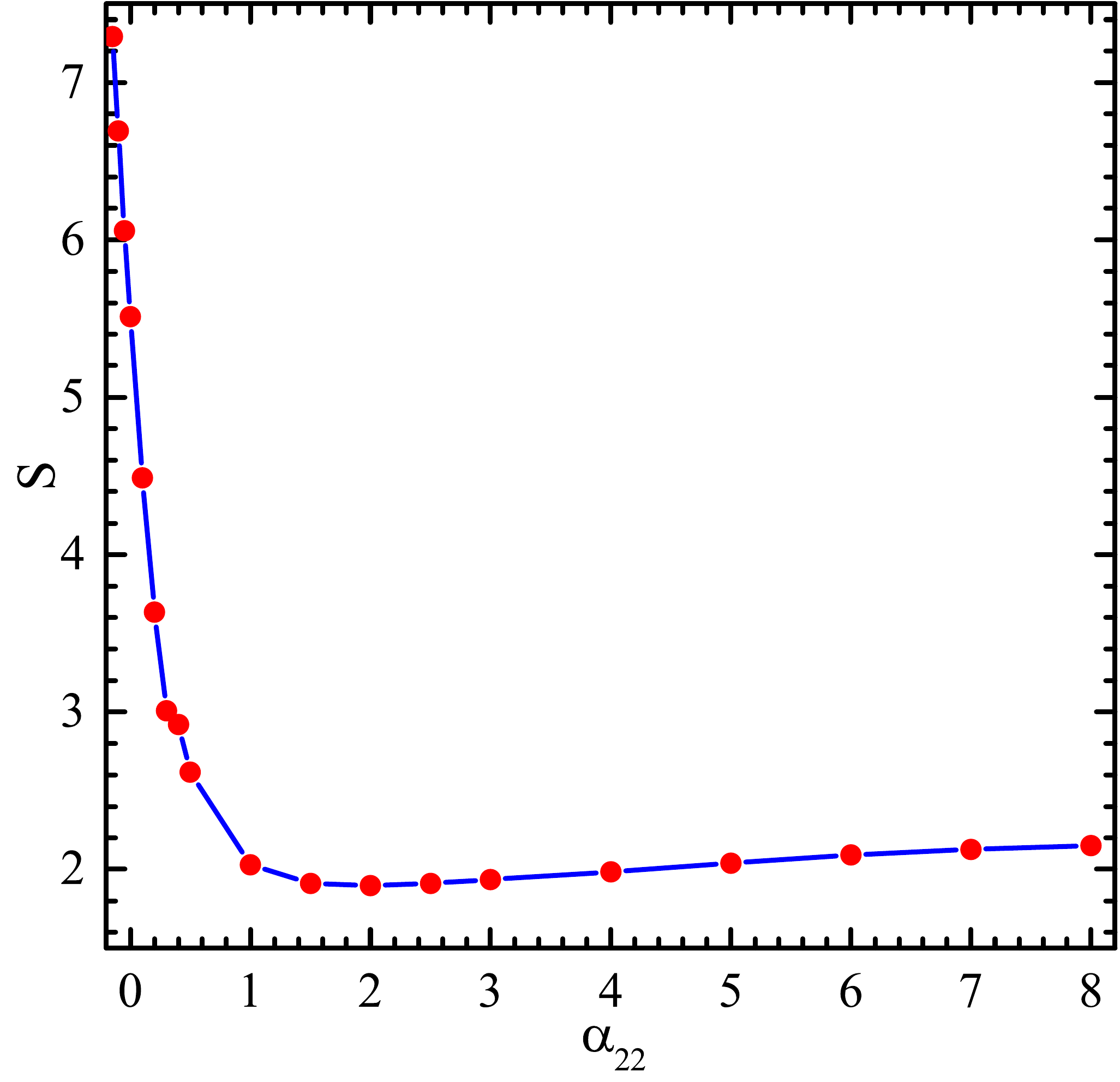}
\hspace{1.0cm}
\includegraphics[type=pdf,ext=.pdf,read=.pdf,width=7.5cm]{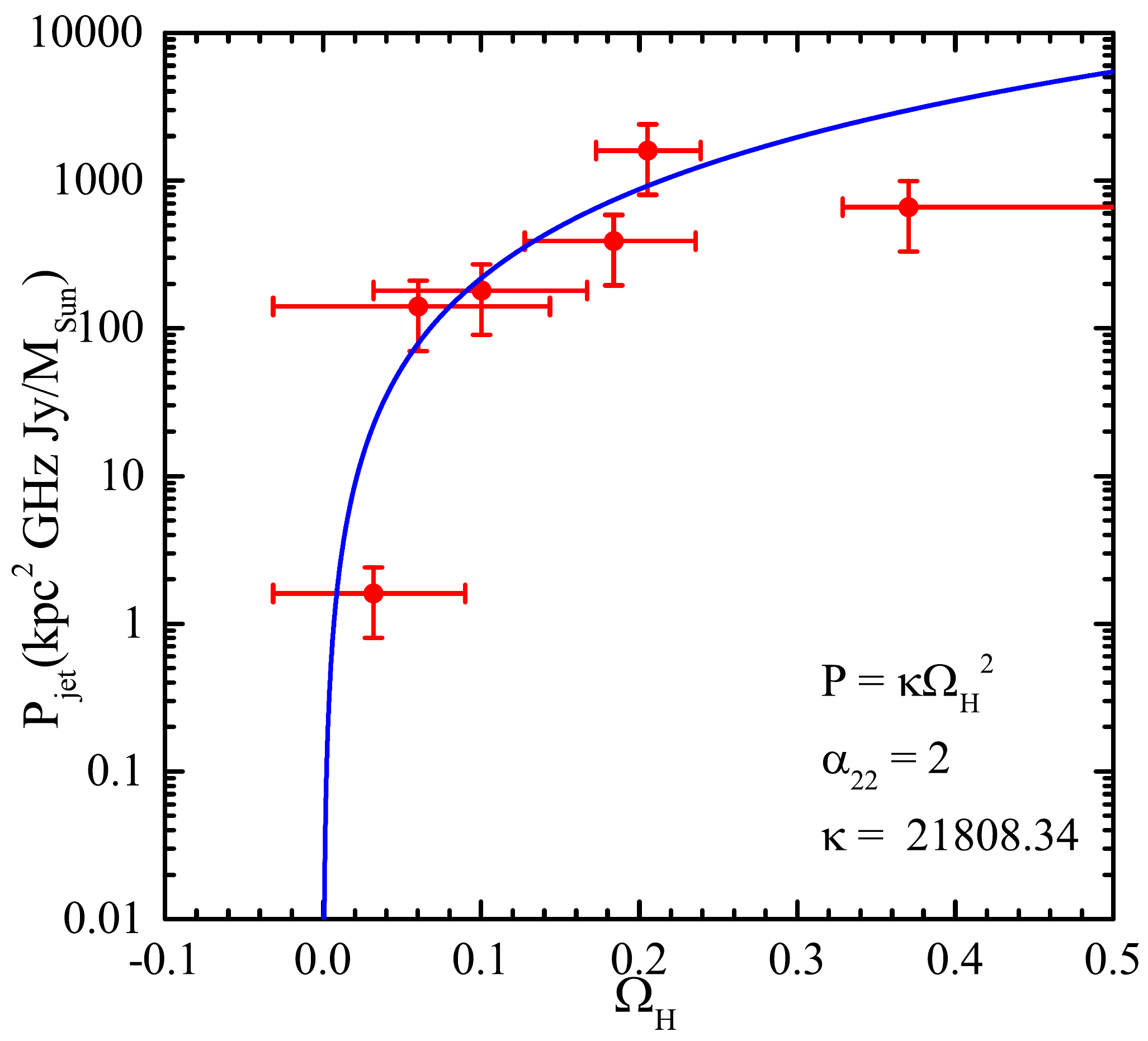} \\
\vspace{0.5cm}
\includegraphics[type=pdf,ext=.pdf,read=.pdf,width=7.5cm]{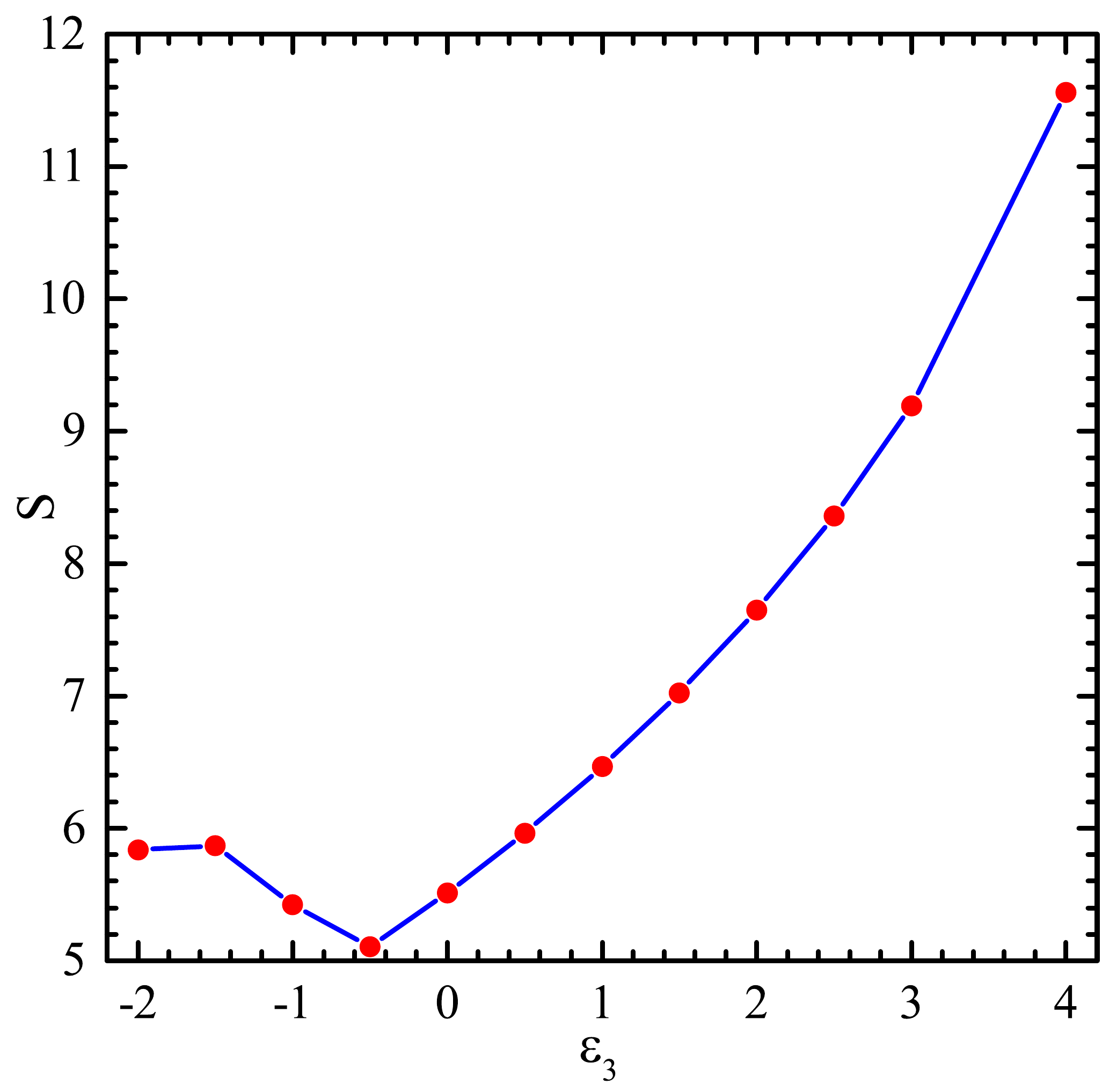}
\hspace{1.0cm}
\includegraphics[type=pdf,ext=.pdf,read=.pdf,width=7.5cm]{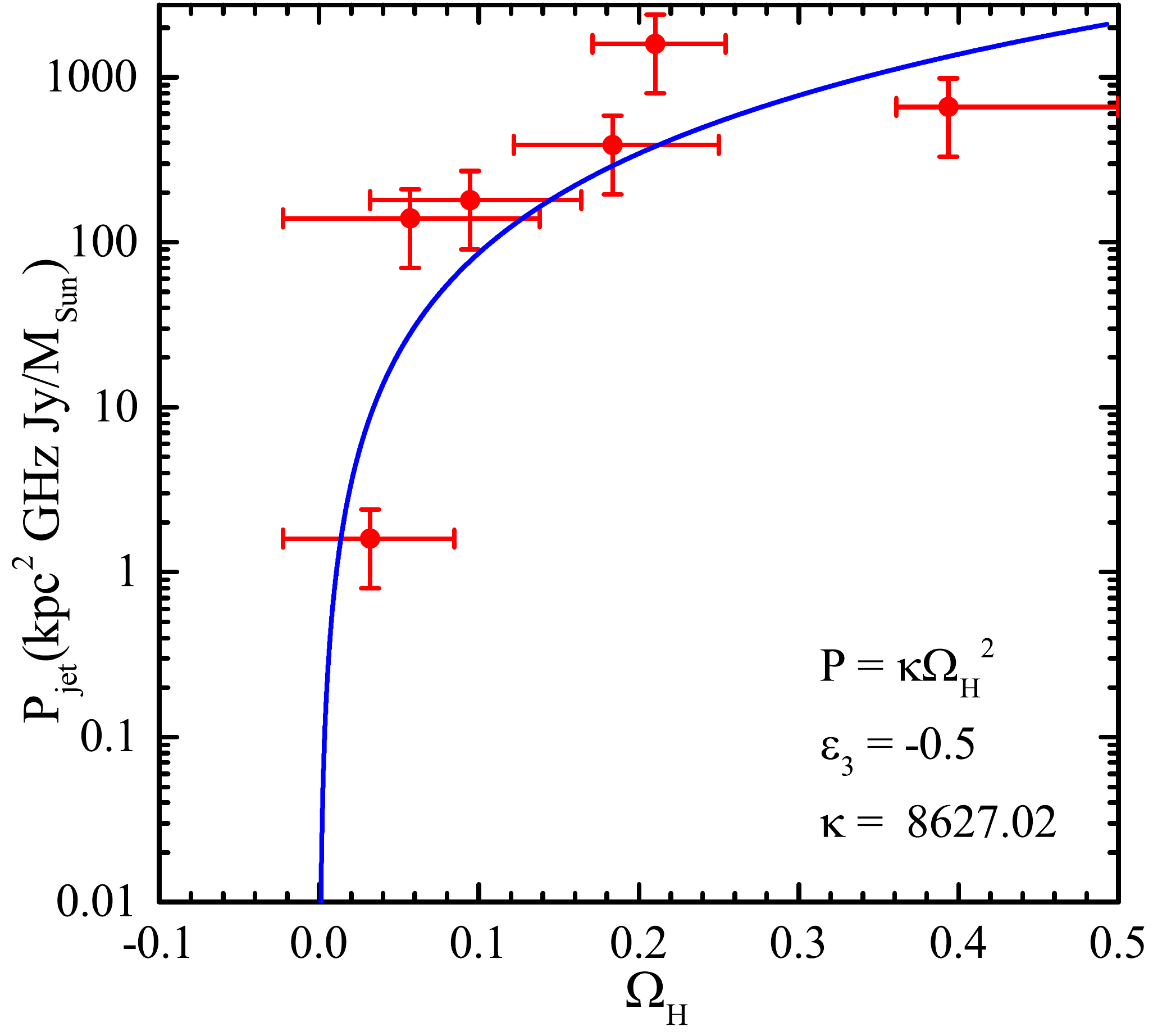} \\
\end{center}
\caption{As in Fig.~\ref{f2G2} but for $\Gamma = 5$. Here the deformation parameters that minimise the $S$ function are: $\alpha_{13}$ = $-0.99$ (top panel), $\alpha_{22}$ = $2$ (middle panel), and $\epsilon_3$ = $-0.5$ (bottom panel). \label{f2G5}}
\end{figure*}

Our results are shown in Figs.~\ref{f2G2} and \ref{f2G5}, respectively for the case $\Gamma = 2$ and $\Gamma = 5$. In both figures, the left panels show $S(i)$ minimised over $k$ as a function of the deformation parameter: $\alpha_{13}$ (top panel), $\alpha_{22}$ (middle panel), and $\epsilon_3$ (bottom panel). The right panels show the data for the parameter where $S$ achieves the minimum value. $S$ is roughly the same as $\chi^2$, so $S_{\rm min} + \Delta S$, where $S_{\rm min}$ is the minimum of $S$ and $\Delta S = 2.30$ and 4.61 correspond, respectively, to the 68\% and 90\% confidence level for two degrees of freedom (we have $k$ and the deformation parameter). From the plots in Figs.~\ref{f2G2} and \ref{f2G5} we cannot put any meaningful constraint on the deformation parameters. This is not surprising since the uncertainties on the measurements are too large.

We investigated the possibility of constraining deviations from the Kerr metric using the data presently available. We found that large uncertainties preclude such a possibility at present. We now take up the question of figuring out the level of accuracy needed to improve the constraints using this approach. We consider the impact on $S$ of two hypothetical sources, measured such that the uncertainties on jet power and spin value (obtained using the Kerr metric assumption) are smaller for the hypothetical sources than for the real sources already considered. Note that we need at least two such sources since $S$ has two degrees of freedom.

\newpage

\begin{figure*}[t]
\begin{center}
\includegraphics[type=pdf,ext=.pdf,read=.pdf,width=7.5cm]{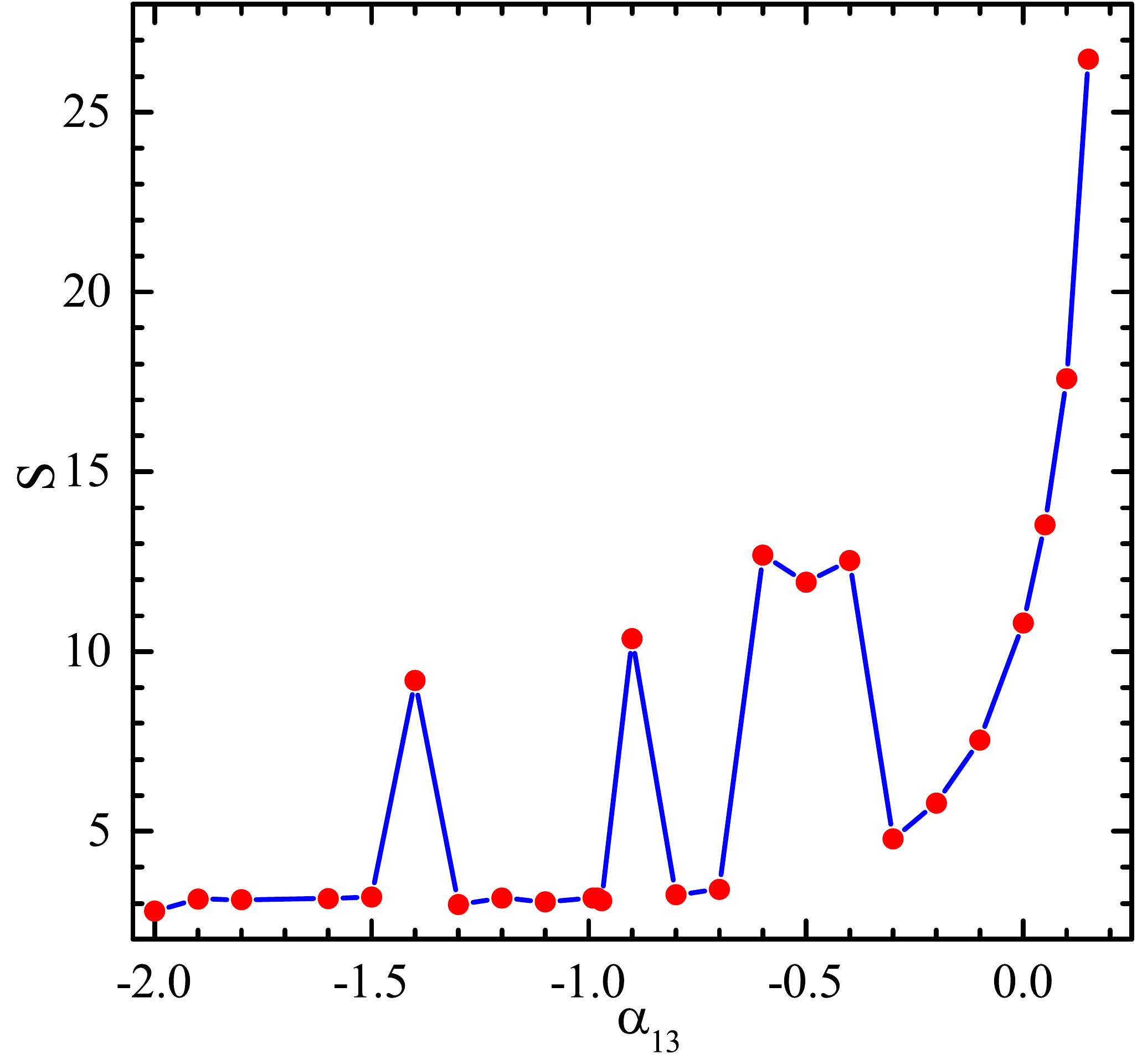}
\hspace{1.0cm}
\includegraphics[type=pdf,ext=.pdf,read=.pdf,width=7.5cm]{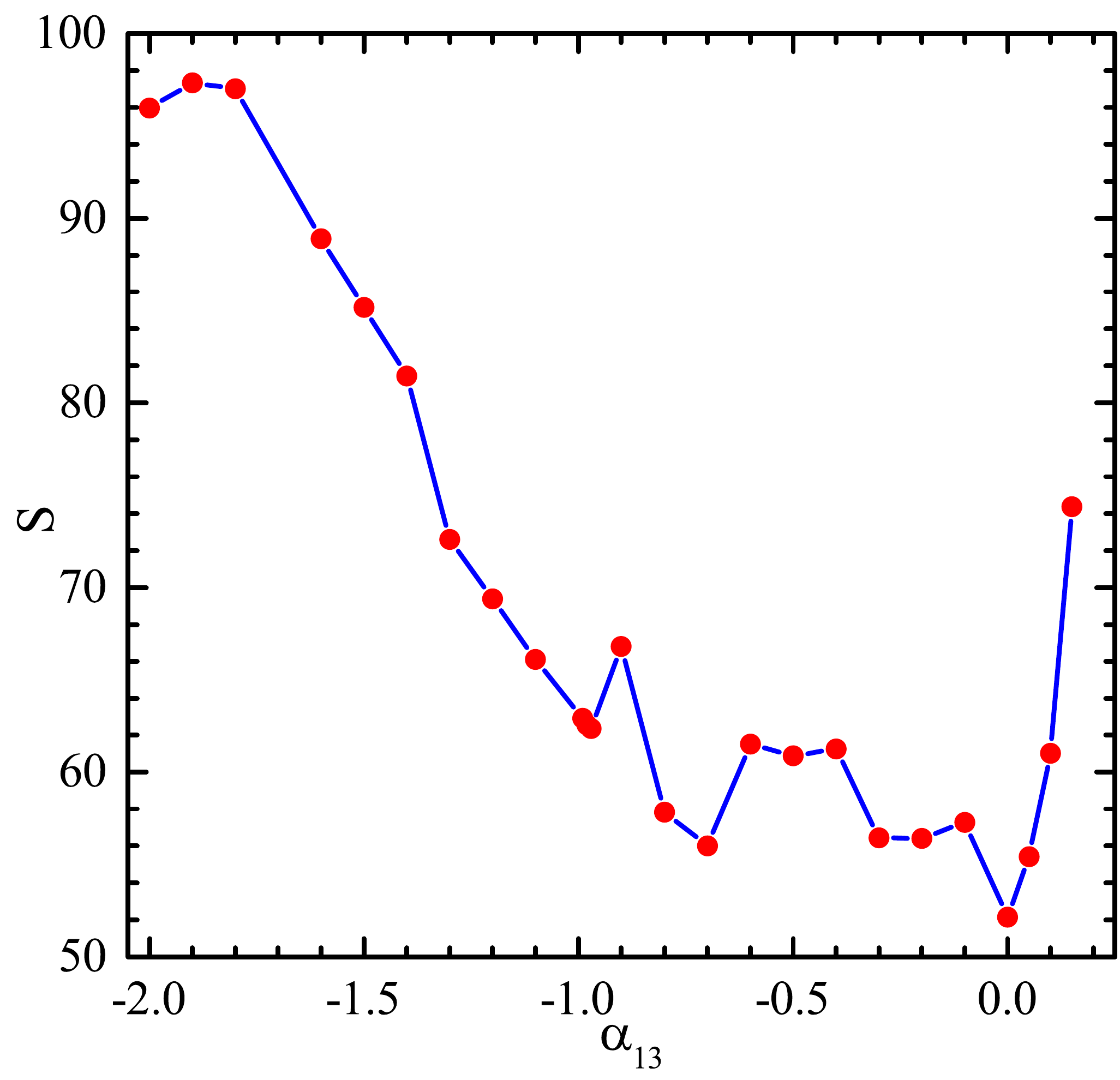} \\
\vspace{0.5cm}
\includegraphics[type=pdf,ext=.pdf,read=.pdf,width=7.5cm]{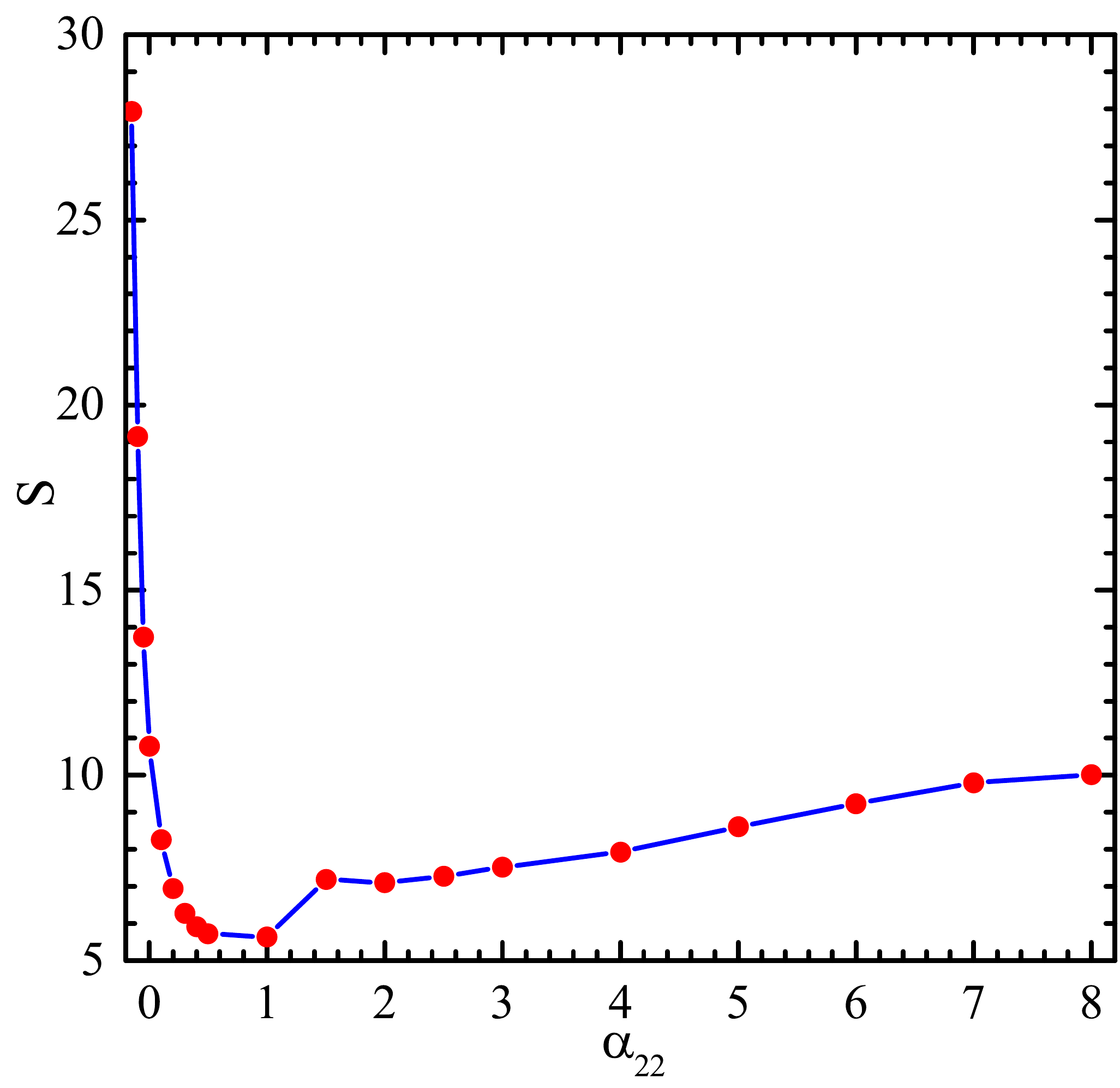}
\hspace{1.0cm}
\includegraphics[type=pdf,ext=.pdf,read=.pdf,width=7.5cm]{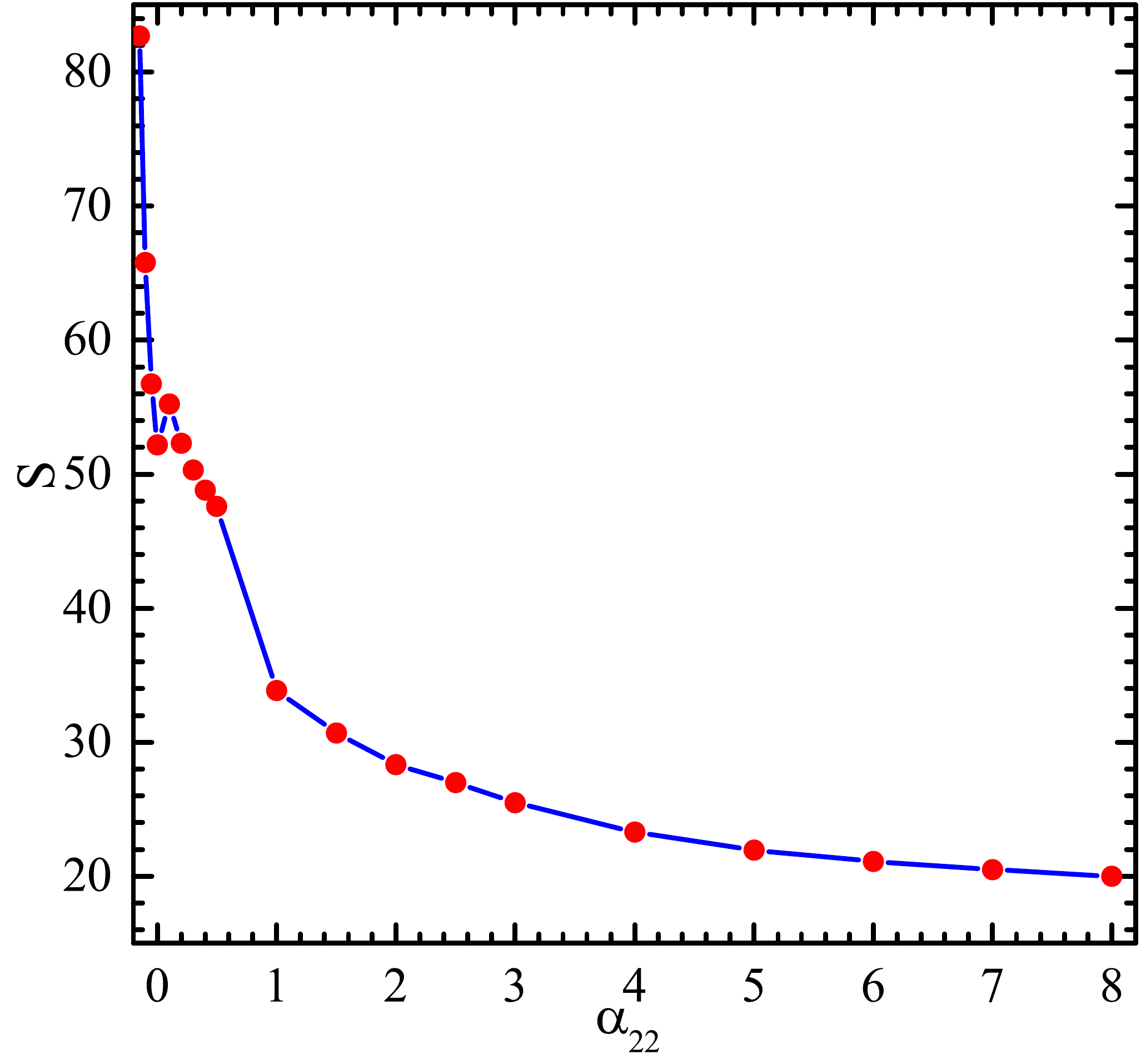} \\
\vspace{0.5cm}
\includegraphics[type=pdf,ext=.pdf,read=.pdf,width=7.5cm]{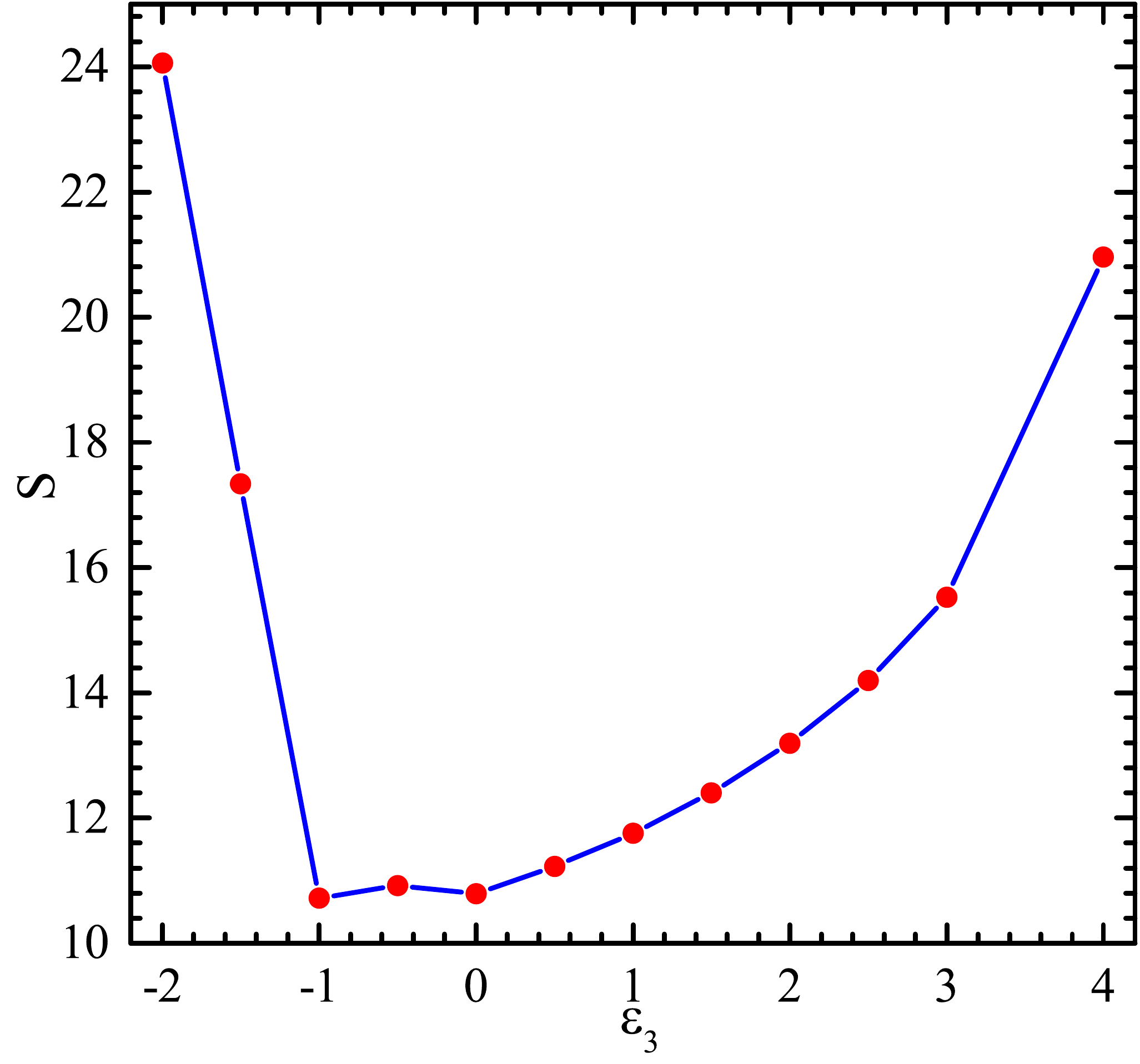}
\hspace{1.0cm}
\includegraphics[type=pdf,ext=.pdf,read=.pdf,width=7.5cm]{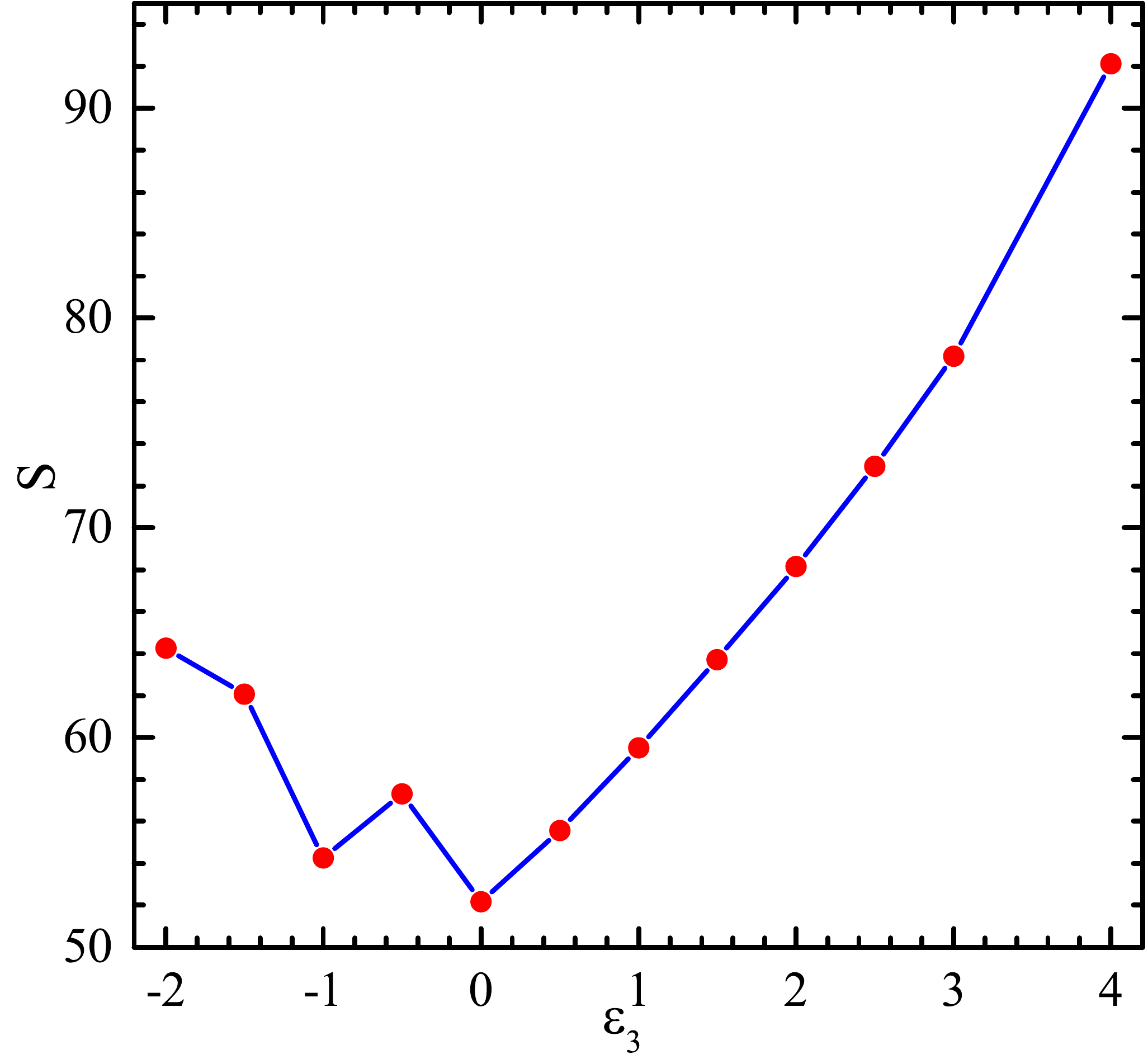} \\
\end{center}
\caption{Left panels: $S$ function of the 6 black hole binaries and one fictitious source minimised over $k$ as a function of the deformation parameters assuming that all jets are produced with $\Gamma = 2$: $\alpha_{13}$ (top panel), $\alpha_{22}$ (middle panel), and $\epsilon_3$ (bottom panel). Right panels: $S$ function of the 6~black hole binaries and two fictitious sources minimised over $k$ as a function of the deformation parameters assuming that all jets are produced with $\Gamma = 2$: $\alpha_{13}$ (top panel), $\alpha_{22}$ (middle panel), and $\epsilon_3$ (bottom panel). \label{f3G2}}
\end{figure*}

\begin{figure*}[t]
\begin{center}
\includegraphics[type=pdf,ext=.pdf,read=.pdf,width=7.5cm]{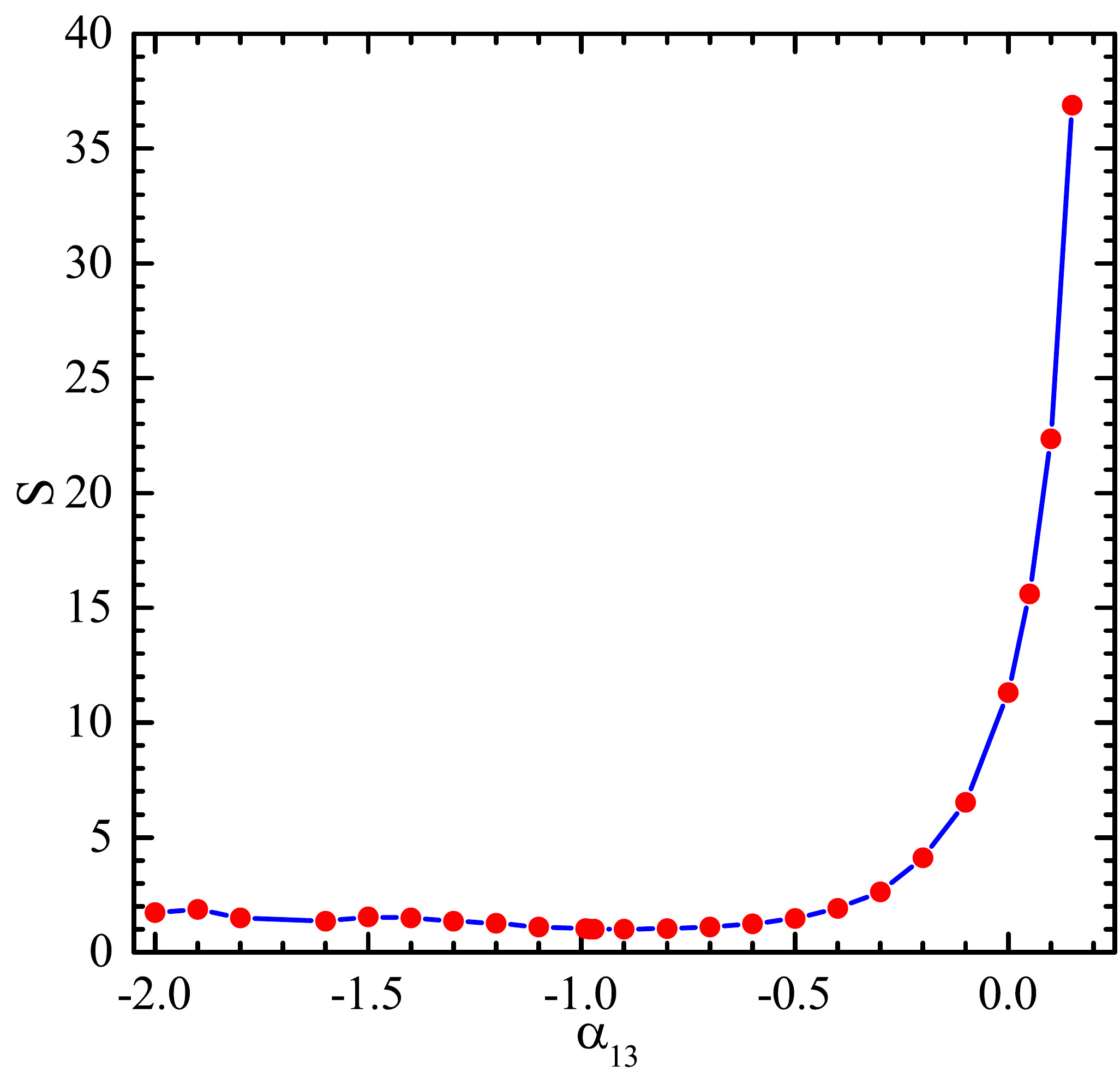}
\hspace{1.0cm}
\includegraphics[type=pdf,ext=.pdf,read=.pdf,width=7.5cm]{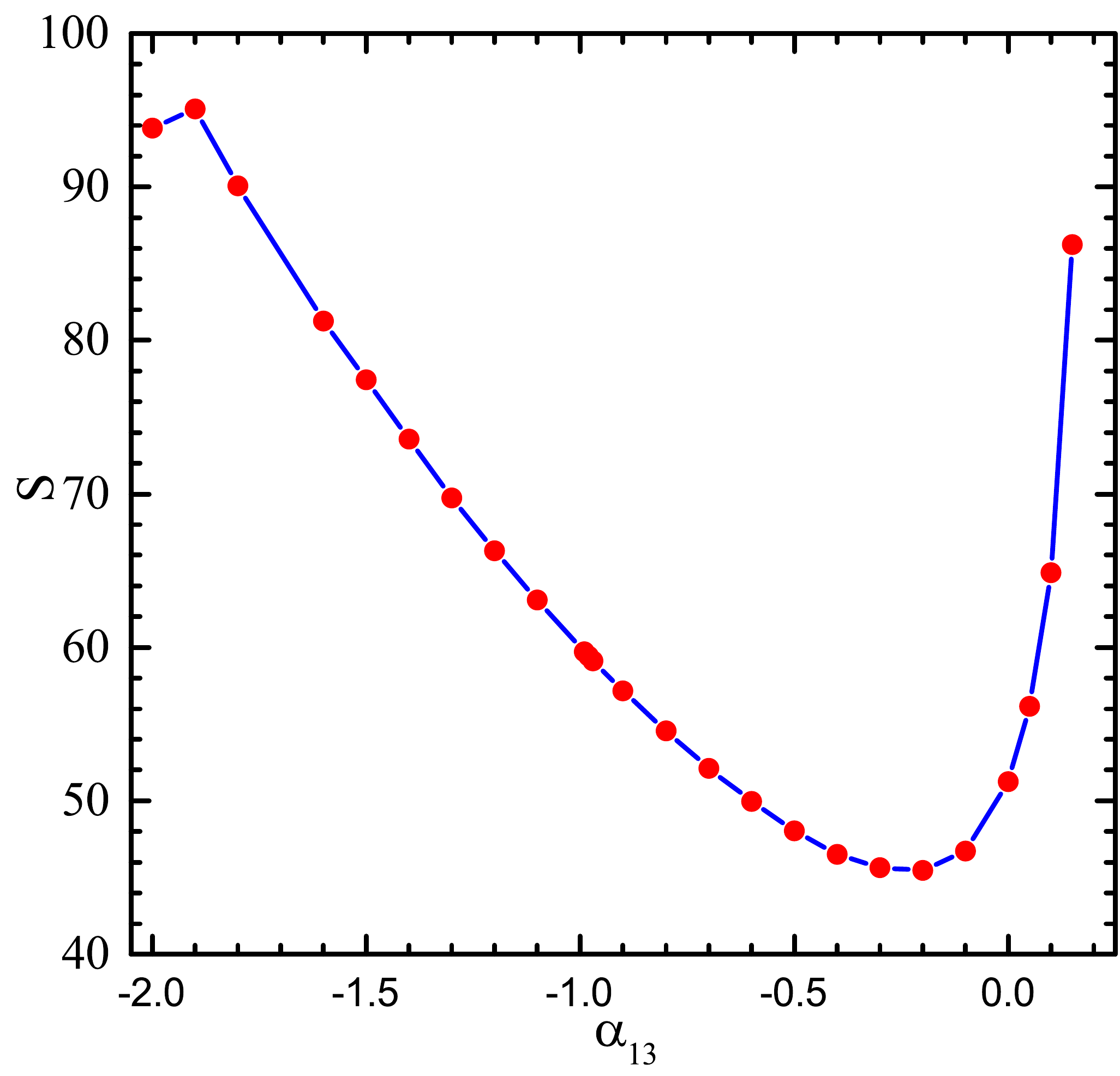} \\
\vspace{0.5cm}
\includegraphics[type=pdf,ext=.pdf,read=.pdf,width=7.5cm]{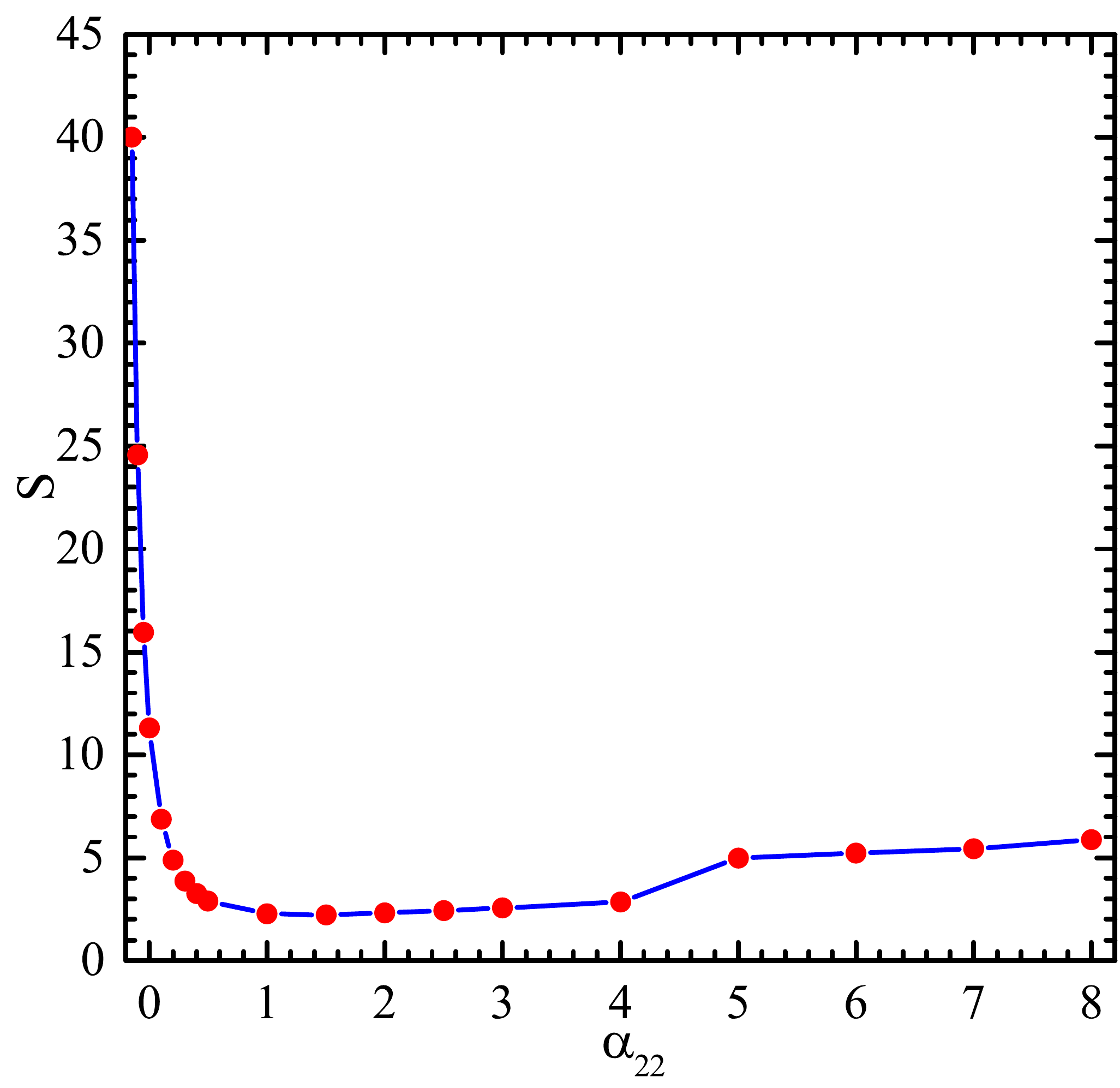}
\hspace{1.0cm}
\includegraphics[type=pdf,ext=.pdf,read=.pdf,width=7.5cm]{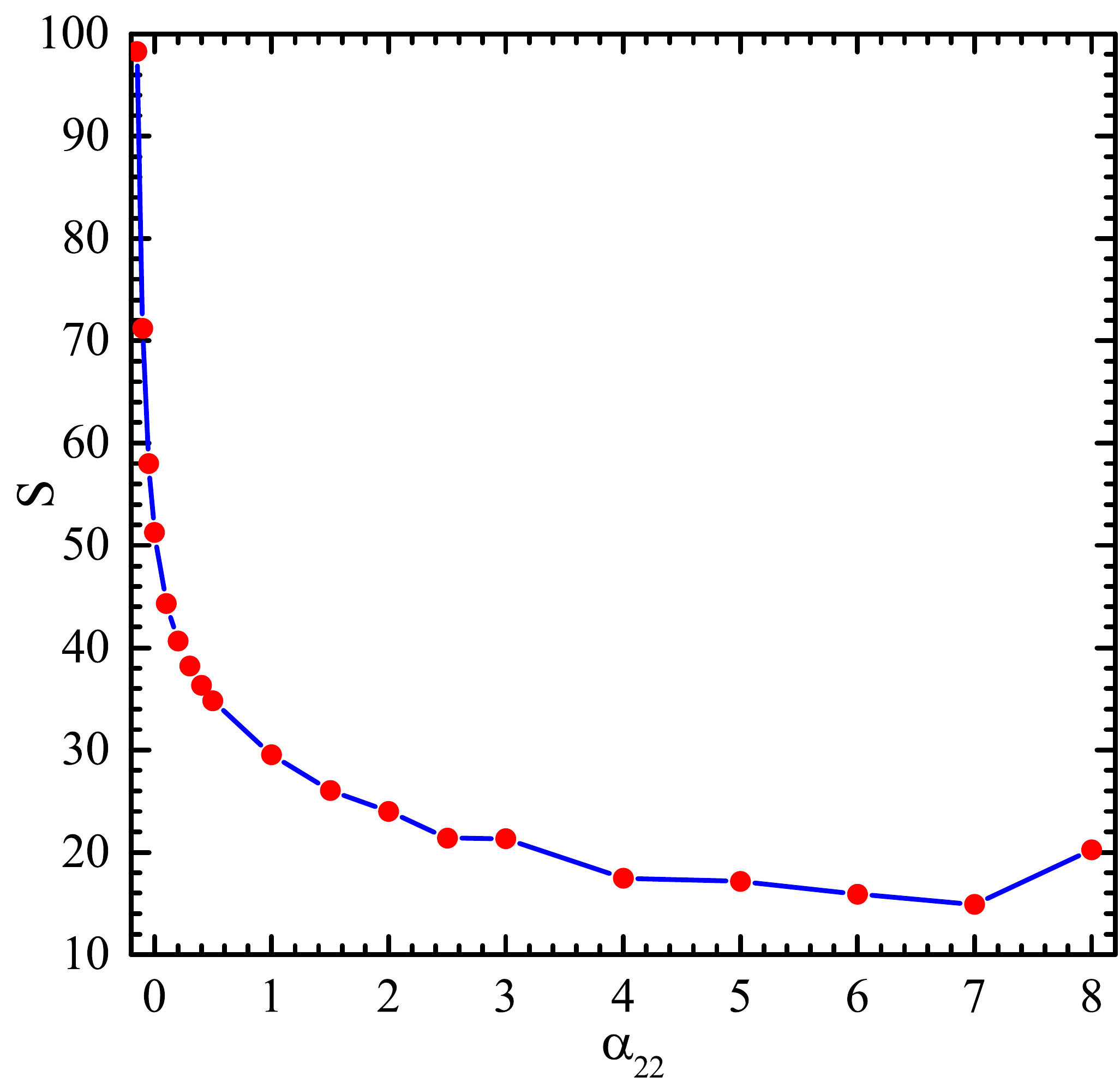} \\
\vspace{0.5cm}
\includegraphics[type=pdf,ext=.pdf,read=.pdf,width=7.5cm]{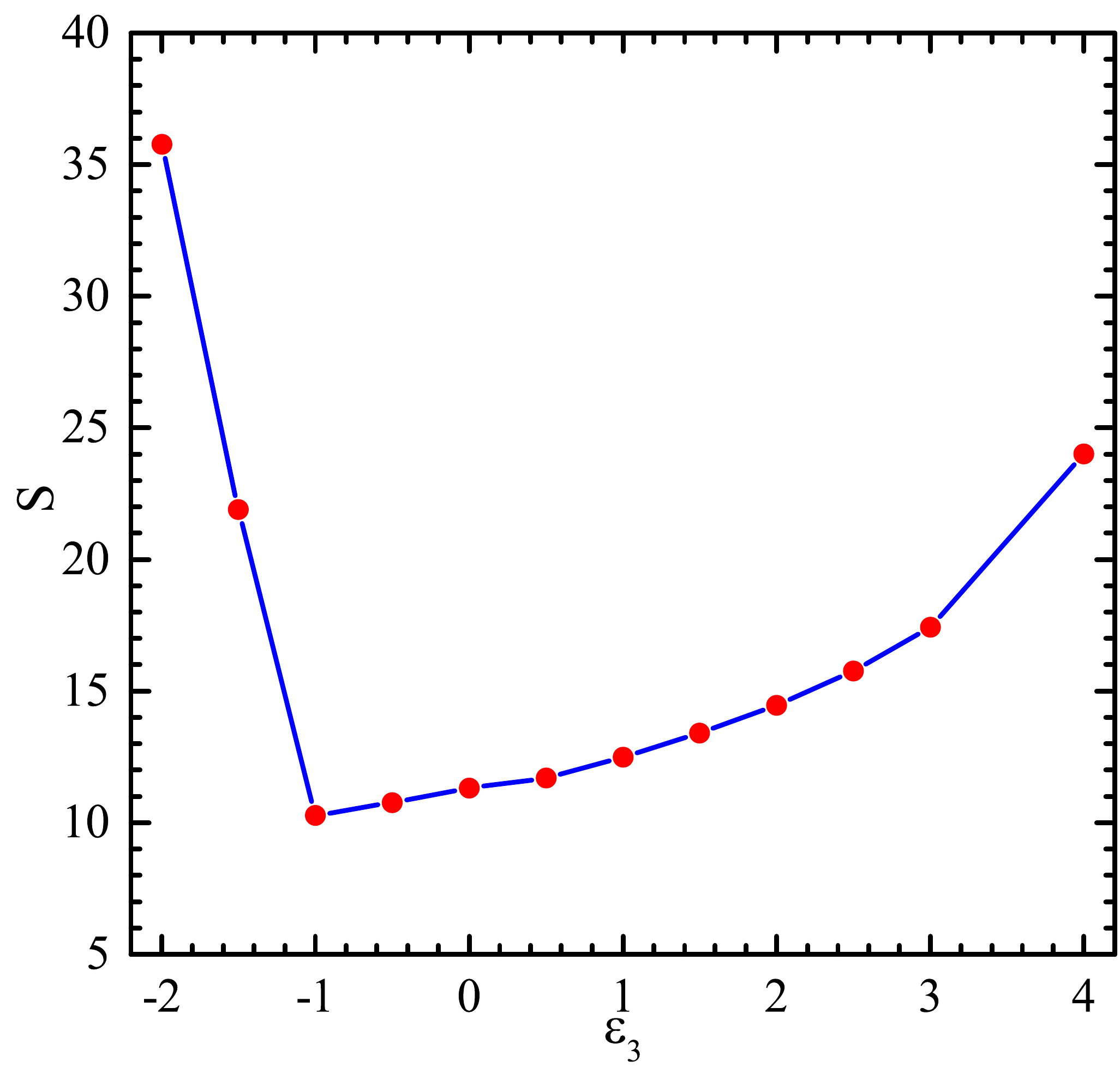}
\hspace{1.0cm}
\includegraphics[type=pdf,ext=.pdf,read=.pdf,width=7.5cm]{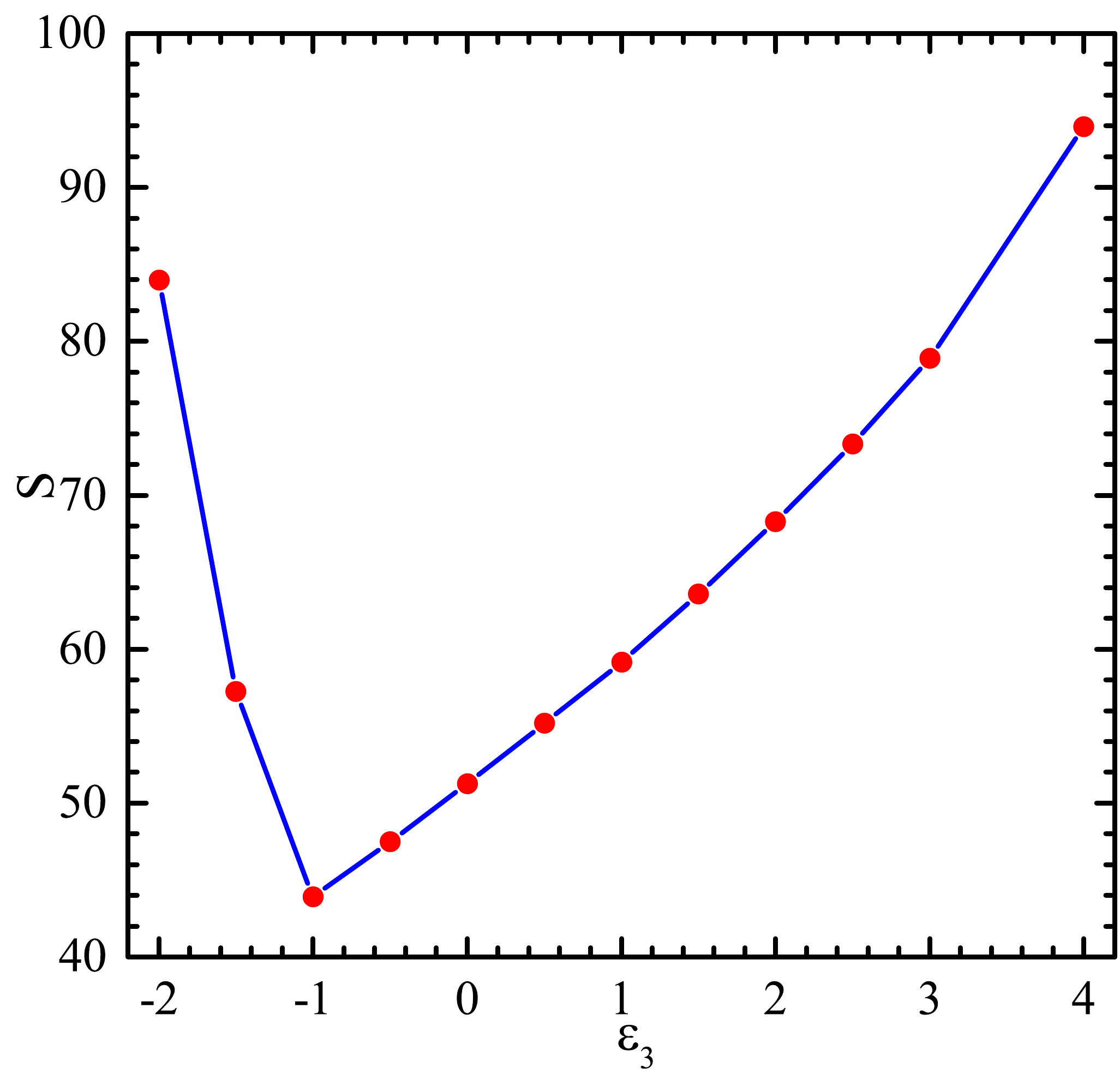} \\
\end{center}
\caption{As in Fig.~\ref{f3G2}  but for $\Gamma = 5$. \label{f3G5}}
\end{figure*}

As the first additional source\footnote{These hypothetical measurements, as well as those in Eqs.~(\ref{eq-spin2a}) and (\ref{eq-spin2b}), are chosen (quite arbitrarily) such that they are sitting on the line for the best fit for the Kerr metric. These hypothetical measurements are only used to get an idea of the necessary precision in the measurement of the spin and the jet power to constrain the deformation parameters, while the values of the parameters in the best fits are not important.}, we have considered the following measurements of spin and jet power:
\be\label{fict1}
a_* &=& 0.90 \pm 0.01 \, , \\
P_{\rm jet} &=&  
  \begin{cases}
    120\pm12 \, , & \quad \Gamma=2 \\
    2700\pm270 \, , & \quad \Gamma=5 \\
  \end{cases}
 \, .
\ee
This would be a black hole with a moderately high spin parameter. The spin measurement has an uncertainty $\Delta a_* = 0.01$, to be compared with $\Delta a_* \approx 0.05$ of today. Such a level of precision is possible, but it would require much better measurements of the black hole mass, black hole distance, and inclination angle of the disk with respect to our line of sight. This will be possible with better optical observations and more robust models to use in the fits of the curve luminosity of the companion star. The estimate of the jet power has an uncertainty of 10\%, to be compared with an uncertainty of 50\% assumed (uncertainties on jet power are difficult to estimate) in the previous cases.
With the 6 real sources and one additional fictitious source with the measurements in Eq.~(\ref{fict1}), the plots in the left panels in Figs.~\ref{f2G2} and \ref{f2G5} become, respectively,  the plots in the left panels in Figs.~\ref{f3G2} and \ref{f3G5}. There is not a significant improvement, and it could not be otherwise, because there are two parameters to fit, the deformation parameter and $k$.

For the second additional source, we assume that the measurements of spin and jet power are
\be\label{eq-spin2a}
a_* &=& 0.30 \pm 0.03 \, , \\
\label{eq-spin2b}
P_{\rm jet} &=&  
  \begin{cases}
    30\pm3 \, , & \quad \Gamma=2 \\
    660\pm66 \, , & \quad \Gamma=5 \\
  \end{cases}
 \, .
\ee
Both the spin parameter $a_*$ and the jet power $P_{\rm jet}$ are measured with a precision of 10\%. The new error functions are shown in the right panels in Fig.~\ref{f3G2} ($\Gamma=2$) and Fig.~\ref{f3G5} ($\Gamma=5$). Having in mind $\Delta S = 2.30$ and 4.61 for 68\% and 90\% confidence levels respectively, we find that much better constraints can be put on the deformation parameters. In the case of $\alpha_{13}$ and $\epsilon_3$, the addition of two fictitious sources (whose measurements are chosen such that they are sitting on the line for the best fit for the Kerr metric) moves the minimum of $S$ to the Kerr solution. This is what we should have expected. For $\alpha_{22}$, we do not see a similar behavior. We do not attribute any particular physical meaning. It is possible that our choice of the values for the additional sources is not perfect, or that there is an intrinsic incompatibility among the existing measurements.

Last, we note that we have considered several simplifications in our analysis. They can be acceptable for an explorative work like our study, but they have to be removed in a more detailed analysis aiming at providing strong constraints on possible deviations for the Kerr metric. In the BZ scenario, the jet power depends on both the angular velocity of the event horizon and the magnetic field strength. Following the argument in Ref.~\cite{n1}, we have not directly taken the magnetic field strength into account, assuming that the power is proportional to the black hole mass. The power extracted from every black hole has been evaluated from the radio emission, neglecting the jet radiative efficiency. This is equivalent to the assumption that the radio emission power is the same fixed fraction of the total power for all the sources. There are large uncertainties on the magnetic field strengths and on the radiative efficiencies, but there have been some progresses recently (see, e.g., Ref.~\cite{kino2015}).

\section{Summary and conclusions}

In the BZ model, black hole jets are powered by the rotational energy of the compact object. In the present paper, we have studied how the BZ model may change in the case of black holes in alternative theories of gravity. As a prototype, we have considered the Johannsen metric, but a similar analysis could be repeated for other non-Kerr black holes. Depending on the non-vanishing deformation parameter in the Johannsen metric, the jet power may or may not be different with respect to that of a Kerr black hole with the same mass and spin. We have also found that some rotating black holes may not generate any jet because; despite of the non-vanishing spin angular momentum, they have no ergoregion.

If the BZ mechanism is responsible for the formation of transient jets in black hole binaries, as suggested in Ref.~\cite{n1}, the measurement of the jet power may be used to probe the geometry of the spacetime and test the nature of the black hole. If the estimate of the jet power is combined with independent measurements of the black hole spins, it is potentially possible to constrain the deformation parameters. We have explored such a possibility by considering the present measurements of the jet powers and of the spins via the continuum-fitting method for the 6~sources for which both measurements are available. Current data cannot yet place any interesting constraints and such an approach cannot currently compete with other techniques that test the Kerr metric with electromagnetic radiation. The point is that current uncertainties in the estimate of the jet power are very large. However, in the presence of much more precise measurements, which might be possible in the future, this approach might be useful at least to constrain those kind of deformations from the Kerr geometry that more significantly affect the angular momentum of the event horizon and might not significantly affect the properties of the radiation emitted from the accretion disk.


\begin{acknowledgments}
CB, SN, and GP were supported by the NSFC (grants 11305038 and U1531117) and the Thousand Young Talents Program. CB also acknowledges support from the Alexander von Humboldt Foundation. MJM appreciates support from an Ernest Rutherford STFC fellowship.
\end{acknowledgments}


\end{document}